\documentclass[%
 reprint,
superscriptaddress,
 amsmath,amssymb,
 aps,
]{revtex4-2}

\usepackage{mathrsfs}
\usepackage{xcolor}
\usepackage{cancel}
\usepackage{graphicx}
\usepackage{dcolumn}
\usepackage{bm}

\begin{document}

\preprint{APS/123-QED}

\title{Imprints of dark matter on the structural properties of minimally deformed
compact stars}

\author{Z. Yousaf}
 \email[Email: ]{zeeshan.math@pu.edu.pk}
\affiliation{Department of Mathematics, University of the Punjab, Quaid-i-Azam Campus, Lahore-54590, Pakistan.}


\author{Kazuharu Bamba}
 \email[Email: ]{bamba@sss.fukushima-u.ac.jp}
\affiliation{Faculty of Symbiotic Systems Science,
Fukushima University, Fukushima 960-1296, Japan.}%

\author{Bander Almutairi}
\email[Email: ]{baalmutairi@ksu.edu.sa}
\affiliation{Department of Mathematics, College of Science,\\
King Saud University, P.O.Box 2455 Riyadh 11451, Saudi Arabia.}%

\author{Yuki Hashimoto}
\email[Email: ]{s2471002@ipc.fukushima-u.ac.jp}
\affiliation{Faculty of Symbiotic Systems Science,
Fukushima University, Fukushima 960-1296, Japan.}%

\author{S. Khan}
\email[Email: ]{suraj.pu.edu.pk@gmail.com}
\affiliation{Department of Mathematics, University of the Punjab, Quaid-i-Azam Campus, Lahore-54590, Pakistan.}

\date{\today}

\begin{abstract}
In this manuscript, we investigate the possibility of constructing
anisotropic dark matter compact stars motivated by the Einasto density
profile. This work develops analytical solutions for an anisotropic
fluid sphere within the framework of the well-known Adler-Finch-Skea
metric. This toy model incorporates an anisotropic fluid
distribution that includes a dark matter component. We use the
minimal geometric deformation scheme within the framework of
gravitational decoupling to incorporate anisotropy into the pressure
profile of the stellar system. In this context, we model the
temporal constituent of the $\Theta$-field sector to characterize
the contribution of dark matter within the gravitational matter
source. We present an alternative approach to studying anisotropic
self-gravitating structures. This approach incorporates additional
field sources arising from gravitational decoupling, which act as
the dark component. We explicitly verify whether the proposed model satisfies all the requirements for describing realistic compact structures in detail. We conclude that the modeling of the Einasto density model with the Adler-Finch-Skea metric
gives rise to the formation of well-behaved and viable astrophysical
results that can be employed to model the dark matter stellar configurations.
\end{abstract}

\maketitle


\section{\label{sec:level1}Introduction}

Understanding the nature of the dark content of our universe remains
largely unexplained, posing a significant challenge for both
theoretical and observational cosmology. The $\Lambda$-cold dark
matter (DM) model suggests that DM constitutes about $27\%$ of the
total cosmic mass density \cite{jarosik2011seven,ade2014planck}, yet
it remains elusive to current astronomical instrumentation. The
initial evidence regarding the existence of DM emerges from
discrepancies between predicted and observed rotation curves of
spiral galaxies \cite{faber1979masses,bosma198121}. Neutralino has
appeared as a primary candidate among the potential explanations for
the principal constituent and origin of DM. It is
classified within the group of the lightest supersymmetric particles
\cite{barger2002indirect,spolyar2008dark}. The investigation by
Hadjimichef \emph{et al.} \cite{hadjimichef2017dark} proposed a
novel DM-based model for modeling ultra-dense stellar
configurations, including white dwarfs, neutron stars (NS), and black
holes. This model considers DM as a standard model fermion gauge
singlets against a dark energy background within the context of
pseudo-complex general relativity (GR). On the other hand, the
mass-radius relationship can be impacted by the presence of DM on
the periphery of NS, as discussed in \cite{lopes2018dark}. Rezaei
\emph{et al.} \cite{rezaei2018neutron,rezaei2018double} used a
polytropic EoS to study how spin-polarized self-interacting DM
affects NS structure. Building on earlier research, scientists
investigated the effects of concentrated DM on the structures of
self-gravitational entities as detailed in
\cite{panotopoulos2017gravitational,li2012gravitational}.
Furthermore, Ciarcelluti \emph{et al.} \cite{ciarcelluti2011have}
explored the potential presence of a DM core within NS.

Analytical closed-form solutions of the field equations of general
relativity (GR) are essential resources for comprehending the strong
gravity regime in ultra-dense stars. Studies show that GR provides a
well-established toolbox of solutions for isotropic objects, relying
on simplified assumptions of spherical symmetry and a staticity.
However, a significant portion of these solutions lack physical
relevance and often fail to pass basic tests based on astrophysical
observations \cite{delgaty1998physical}. On the other hand,
incorporating anisotropic matter content as an approximation to
study stellar systems has consistently garnered attention and
remains a dynamically evolving area of study in astrophysical
scenarios. Herrera \emph{et al.} \cite{herrera1997local} discussed
various theoretical lines of evidence suggesting that diverse and
intriguing physical phenomena can induce a significant prevalence of
local anisotropies across low and high-density regimes.
Supplementing the analysis of high-density regimes, Ruderman's
investigations into the formation of more realistic stellar objects
offer insights into the potential anisotropy of nuclear matter
\cite{ruderman1972pulsars}. According to these studies, nuclear
matter may exhibit anisotropic behavior, especially at high-density
levels $\thicksim3\times10^{17} \text{kg/m}^3$. In these regimes, a
relativistic treatment of nuclear interactions becomes significant.
The origin of anisotropic behavior is inherently linked to the
deviation from the standard pressure term within the stress-energy
tensor (SET). In an isotropic self-gravitational distribution, the
radial pressure ($P_{r}$) and tangential pressure ($P_{t}$) are
equal: $P_{r}=P_{t}$.  However, stellar fluids featuring anisotropic
matter content are characterized by a SET where $P_{r} \neq P_{t}$.
To enhance the understanding of gravitational interactions and the
formation of stellar structures for a broader context, it is
valuable to acknowledge the studies presented in
\cite{poplawski2013intrinsic,poplawski2021gravitational,poplawski2012four,poplawski2013cosmological,khlopov1985gravitational,
di2017spin,de2020general,de2021testing,astashenok2020supermassive,nojiri2020f}

Building upon the seminal research conducted by Bowers and Liang \cite{bowers1974anisotropic}, a significant amount of attempts have been dedicated to further elucidating the behavior and characteristics of astrophysical entities filled with anisotropic matter content. Additionally, the incorporation of matter configurations featuring anisotropic content has been extensively probed in the study of stellar structures \cite{heintzmann1975neutron,cosenza1982evolution,herrera1985isotropic,chan1993dynamical,
di1997cracking,dev2002anisotropic,yousaf2022f,malik2024charged,malik2024singularity}. In this direction, the research conducted by Lake and Delgaty \cite{delgaty1998physical} represented a significant advancement in our understanding of stellar interiors. Through the utilization of the spherically symmetric metric ansatz with isotropic fluid, their investigation revealed that only a small fraction (9 out of 127) of known isotropic solutions were physically relevant. This significant constraint triggered a surge in research, focusing on unraveling the fresh insights brought about by local anisotropies within compact configurations.
Anisotropic behavior in self-gravitational stellar fluids can be attributed to several physical mechanisms, including viscosity, rotation, the presence of superfluid, magnetic field, or the existence of a solid core (see \cite{herrera1997local} and references therein). Anisotropies can also be induced by certain phenomena like phase transitions or pion condensation, among other contributing factors \cite{sokolov1980phase,sawyer1972condensed}. Understanding the behavior of astrophysical dark configurations under an anisotropic framework is a well-studied topic in cosmology and astrophysics. This pressure anisotropy holds particular significance in the analysis of ultra-compact entities such as black holes, NS, wormholes, and other unique stellar formations \cite{albalahi2024electromagnetic,albalahi2024isotropization,yousaf2024modeling}. Researchers have investigated these effects using both standard four-dimensional metric ansatzes \cite{maurya2016relativistic} and within the framework of higher-dimensional braneworld scenarios \cite{germani2001stars,ovalle2008searching,ovalle2015brane}.

The inclusion of anisotropy as a fluid approximation in the modeling
of collapsing stellar structures such as neutron stars or quark
stars give rise to several interesting and significant features: i)
it increases the stability of the stellar fluid; ii) it enables the
construction of highly compact stellar objects; iii) the
gravitational surface redshift achieves higher magnitudes compared
to its isotropic equivalent; iv) the inclusion of an additional term
increases the hydrostatic equilibrium, which acts to counterbalance
gravitational collapse into a point singularity provided that the
anisotropic scalar $\Pi$, remains positive throughout. Despite the
increasing interest in local anisotropies, there are still several
significant challenges in effectively including them into
distributions of stellar matter. Here are some essential questions
that require attention:
\begin{enumerate}
  \item What are the efficient approaches for including the effects of localized pressure variations (anisotropies) for describing the interior of compact configurations?
  \item Does the feasibility of the resultant solution rely on the particular method employed to incorporate anisotropy?
  \item Do the numerical results obtained from anisotropic models align with experimental data from observations of real stellar structures?
\end{enumerate}
An EMT with varying pressure in the principal directions of the
stellar fluid or the inclusion of an electromagnetic field are the
most straightforward approaches to generate anisotropies into the
self-gravitational objects. Interestingly, the approach of
decoupling the spherical stellar distribution using the minimal
geometric deformation (MGD), introduced within the framework of GR
emerges as a natural means to introduce anisotropies in compact
stars \cite{ovalle2017decoupling,ovalle2018anisotropic}. Originating
in Randall-Sundrum braneworld scenarios
\cite{casadio2015minimal,ovalle2016extending,ovalle2009nonuniform,
ovalle2010schwarzschild,casadio2012brane,ovalle2013tolman,ovalle2013role,casadio2016stability,da2017dark},
the notion of MGD decoupling has a long and well-established
history. This scheme has also found applications in constructing
black hole solutions by using the minimal deformation (radial metric
deformation) in the classical Schwarzschild metric. Following its
success in different frameworks, the MGD scheme was successfully
adapted for use within the arena of GR and enables the researchers
to anisotropize the isotropic stellar models
\cite{ovalle2017decoupling,ovalle2018anisotropic}. As we will
explore in detail in the next section, the MGD approach relies on
two fundamental components: i) the system includes two independent
gravitational-field sources terms represented by $T_{\mu\nu}$ and
$\Theta_{\mu\nu}$; ii) The MGD approach specifically deforms the
radial metric potential through a geometric deformation. This
deformation enables the decoupling of the anisotropic gravitational
source into two sets of differential equations, each corresponding
to a distinct field source. Importantly, the structure of
gravitational decoupling usually considers $\widetilde{T}_{\mu\nu}$
as a known isotropic matter content. The function of
additional-field source $\Theta_{\mu\nu}$ is therefore to insert
localized anisotropies within the considered system. Consequently,
this framework can be regarded as a means of extending isotropic
solutions to anisotropic domains. Here, both the particular
geometric deformation and the constituents of the
$\Theta_{\mu\nu}$-field source are initially unknown. The primary objective becomes establishing additional constraints to determine
them. Numerous strategies have been suggested to tackle this issue,
such as mimic constraint (density-like and pressure-like
constraints), anisotropic mechanism, or imposing a well-behaved
decoupling function.

The MGD-based decoupling approach has shown remarkable versatility,
with a significant increase in its applications across various
scenarios
\cite{fernandes2018gregory,contreras2018minimal,panotopoulos2018minimal,acontreras2018minimal,
gabbanelli2019causal,hensh2019anisotropic,torres2019anisotropic}.
Significantly, progress has been achieved in both the inverse
problem and its extended counterpart
\cite{ovalle2019decoupling,zcontreras2018minimal}. In light of the
outstanding adaptability that the MGD decoupling offers, the onset
of the $\Theta_{\mu\nu}$-field source remains an open question. In
this direction, the authors of
\cite{ovalle2018black,tello2020minimally} analyzed the intriguing
possibility of interpreting $\Theta_{\mu\nu}$ as DM, while its
contribution within the cosmological scenarios, such as cold DM and
$\Lambda$ cold DM model has been presented in
\cite{cedeno2020gravitational}. Motivated by the previous research,
this study explores the possible existence of DM stellar structures
within the arena of GR by employing MGD decoupling. Choosing
suitable parameters that align with observational data ensure that
the resulting DM solutions are physically realistic and
mathematically sound. Our approach considers that the temporal
component of the $\Theta$-field source mimics the Einasto DM
profile. Consequently, the deformation function, $f(r)$ incorporates
this DM contribution to the minimally deformed Adler-Finch-Skea
metric and the remaining components of the SET. The mimicking of the
Einasto DM profile with the temporal component of $\Theta$-sector
makes this work different from the previous studies
\cite{tello2020minimally,maurya2024compact,maurya2023minimally} that
consider the isothermal DM profile for modeling stellar structures.
The Einasto density model has been proven significant in modeling
the compact configurations emersed DM haloes
\cite{yousaf2023generating}.

{In this work, we have examined the influence of DM on the structural features of a spherically symmetric, anisotropic stellar configuration by employing the MGD version of gravitational decoupling. The Einasto profile, which is a popular framework for modeling DM halos in galaxies, is employed in our model to describe the DM component. Although the existence of DM may raise questions about the validity of the vacuum exterior, we want to emphasize that our DM profile is limited to the stellar interior and does not extend significantly beyond the stellar radius. The DM distribution is significantly compressed by the strong gravitational potential of the star core, even though this profile typically extends to large radii. The dense surroundings inside the star serve as a gravitational trap, confining the DM particles and preventing them from leaving. This means that most of the mass of DM is contained within the star radius, and only small amounts are found outside of it. Therefore, the impact of DM on the exterior metric is negligible, validating the use of the vacuum Schwarzschild metric. Several observational studies have investigated the distribution of dark matter (DM) within stellar interiors in recent years. The authors of \cite{ivanytskyi2020neutron} explore the impact of DM on the mass-radius relationship of neutron stars (NS). They show that the observable characteristics of NS can be affected by the presence of DM, providing indirect evidence for the confinement of DM within the stellar interior. Furthermore, the studies discussed in \cite{freese2008stellar} examine the implications of DM annihilation on the stellar interiors and indicate how DM particles are confined within the core of stellar distributions.}

{Lavallaz and Fairbairn \cite{de2010neutron} studied the accretion of DM onto the internal features of NS. This accretion mechanism supports the confinement of DM within stars. Freese \emph{et al.} \cite{raen2021effects} probed potential effects of DM on the evolution of self-gravitational compact stars. They investigated how DM particles confined within stellar configurations can affect their physical characteristics, suggesting the presence of DM in stellar interiors. Furthermore, the increased mass concentration has been already examined in \cite{chakraborty2018packing} by taking into account that the Kalb-Ramond field influences the matter field. A primary difference between this scheme and the present investigation is the inclusion of the Kalb-Ramond field, which influences the exterior metric. This contrasts with the procedure of MGD-decoupling, where the contributions of extra-field source $\Theta_{\mu\nu}$ are neglected \cite{ovalle2018anisotropic}. Thus, the outer manifold remains characterized by a vacuum spacetime. In this respect, Tello-Ortiz \cite{tello2020minimally} constructed minimally deformed DM stellar configurations based on the GD scheme by matching the interior metric with the vacuum Schwarzschild metric. On the other hand, Maurya \emph{et al.} \cite{maurya2023minimally} presented some viable models describing the spherically symmetric MGD-based, DM stelar structures by using the pseudo-isothermal density (PID) profile. Furthermore, the evolution of relativistic distributions with DM induced anisotropy based on the PID profile has been discussed in \cite{maurya2024compact}. These studies were based on the matching of the interior peacetime with the outer Schwarzschild metric. Moreover, several other investigations justify the matching of the interior metric with exterior vacuum geometry by suppressing the contributions of the $\Theta_{\mu\nu}$ sector \cite{sharif2023effects,sharif2023charge,sharif2023effect,yousaf2024role}.}

The remainder of this paper is
structured as follows: The relevance of the Einasto density model in
modeling the DM haloes within the galactic structures along with the
key concepts are presented in Sec. \textbf{II}. In Sec.
\textbf{III}, we provide a concise overview of the Class I approach
and the notion of gravitational decoupling utilizing minimal
geometric deformation. Section \textbf{IV} covers the investigation
of anisotropic stellar solutions endowed with spherical symmetry by
employing the familiar Adler-Finch–Skea metric and Einasto density
profiles through the MGD approach. Section \textbf{V} captures the
influence of the decoupling parameter $\chi$ on the thermodynamical
features of the minimally deformed DM Adler-Finch-Skea metric.
Finally, the last section describes the main findings of this study.

\subsection{\label{sec:level2}Einasto Dark Matter Density Profile}

The Einasto density model for DM distribution, originally proposed by Einasto to describe the galactic halos of M31, M32, M87, and the Milky Way \cite{einasto1965construction}, represents a spherically symmetric density distribution. It is characterized by the following expression  \cite{retana2012analytical}
\begin{align}\label{s32}
\gamma(r)\equiv-\frac{d\ln\rho(r)}{d\ln r}\propto r^{1/n}.
\end{align}
Here, $\rho$ denotes the energy density and $n$ is the shape parameter. The integration of Eq. \ref{s32} provides
\begin{align}\label{s333}
\rho(r)=\rho_{s}\exp\left\{-d_{n}\left[\left(\frac{r}{r_{s}}\right)^{1/n}-1\right]\right\},
\end{align}
\begin{align}\label{s34}
\rho(r)=\rho_{0}e^{-\left(\frac{r}{h}\right)^{1/n}},
\end{align}
where $h$ is the scale length and $\rho_{0}$ is the central density defined as, $h=r_{s}/d^{n}_{n}$ and
$\rho_{0}=\rho_{s}e^{d_{n}}$.
\begin{align}\label{s35}
M=\int^{\infty}_{0}4\pi\rho(r)u^{2}du=4\pi\rho_{0}h^{3}n\Gamma(3n).
\end{align}
The central density can be defined in terms of the mass function $M$ as
\begin{align}\label{s36}
\rho(r)=\frac{M}{4\pi h^{3}n\Gamma(3n)}e^{-\left(\frac{r}{h}\right)^{1/n}}.
\end{align}
Einasto \cite{einasto1969andromeda,einasto1969galactic} pointed out the significance of model descriptive functions satisfying specific constraints to ensure realistic galactic system models. In constructing such models, the density function $\rho(r)$ is the most practical foundation, owing to its mathematical relationship with other key galactic properties. Specifically, the cumulative mass profile, gravitational potential, and surface mass density can all be derived from $\rho(r)$ through integral transformations, highlighting the central role of the $\rho(r)$ in galactic modeling.
Additionally, a physically viable model must have several properties:
  (i) $0<\rho(r)<\infty$,~$\forall~r>0$;
  (ii) $\rho(r)$ should be a smoothly decreasing function that asymptotically approaches zero at large radii;
  (iii) the descriptive functions must not exhibit jump discontinuities;
  and (iv) certain moments of the mass function must be finite, particularly those that define the total mass, the effective radius, and the central gravitational potential of the system.
The galactic models proposed by Einasto used were multi-component in structure. Each component in these models represents a separate, physically uniform stellar population distinguished by its set of parameters $\{\rho_{0},h,n\}$. By tuning the triplet of parameters $\{\rho_{0},h,n\}$, a diverse range of astrophysical objects can be modeled. For example,
\begin{itemize}
  \item $4.54\precsim n\precsim 8.33$ for DM haloes spanning masses from dwarf to cluster scales \cite{navarro2004inner}.
  \item $n\thicksim5.88$ for galaxy-sized haloes \cite{hayashi2008understanding}.
  \item $n\thicksim4.35$ for cluster-sized haloes in the Millennium Run simulation \cite{springel2005simulations,gao2008redshift}.
  \item $n\thicksim3.33$ for the most massive haloes of Millennium Run \cite{springel2005simulations}.
\end{itemize}
The ability to adjust parameters provides flexibility for modeling a range of astrophysical systems.

\section{\label{sec:level3}Class I and MGD Approach to Gravitational Decoupling}

A metric is classified as Class I, indicating it can be embedded in a 5-dimensional pseudo-Euclidean space, when there exists a symmetric tensor $K_{\mu\nu}$ serving as the second fundamental form, which satisfies the Gauss–Codazzi equations
\begin{align}\label{a1}
&R_{\mu\nu\sigma\epsilon}=\varepsilon\left(K_{\mu\sigma}K_{\nu\epsilon}-K_{\mu\epsilon}K_{\nu\sigma}\right),
\\\label{a2}
&\nabla_{\sigma}K_{\mu\nu}=\nabla_{\nu}K_{\mu\sigma},
\end{align}
where $\varepsilon$ can be positive or negative depending on whether the normal vector to the manifold is timelike (-) or spacelike (+). Furthermore, $R_{\mu\nu\sigma\epsilon}$ and $\nabla_{\mu}$ denote the Riemann tensor and covariant derivative, respectively. The line element corresponding to the spherically symmetric fluid sphere in the Schwarzschild-like coordinates, $x^{\mu}\equiv x^{0,1,2,3}=(t,r,\theta,\phi)$, reads
\begin{align}\label{a3}
ds^{2}=e^{\alpha}dt^{2}-e^{\beta}dr^{2}-r^{2}d\Omega^{2},
\end{align}
where $d\Omega^{2}=dr^{2}-\sin^{2}\theta d\theta^{2}$. To maintain the static nature of the aforementioned metric, both $\alpha$ and $\beta$ must be functions of the radial variable $r$ only. Additionally, when taking these factors into account, the timelike four-velocity adopts the following form
\begin{align}\label{a4}
\xi^{\mu}=(e^{-\alpha/2},0,0,0) \quad \textmd{such that} \quad \xi^{t}\xi_{t}=1.
\end{align}
Due to spherical symmetric, the non-null components associated with the tensor $K_{\mu\nu}$ are: $K_{tt}$, $K_{rr}$, and $K_{\theta\theta}=\sin^{2}\theta K_{\phi\phi}$. Substituting these components in Eq. \eqref{a1}, we obtain
\begin{align}\label{a6}
R_{rtrt}=\frac{R_{r\theta r\theta}R_{\phi t\phi t}+R_{r\theta\theta t}R_{r\phi r\phi}}{R_{\theta\phi\theta\phi}},
\end{align}
which gives rise to the following differential equation
\begin{align}\label{a7}
2\frac{\alpha''}{\alpha'}+\alpha'=\frac{\beta'e^{\beta}}{e^{\beta}-1}, \quad \textmd{such that} \quad e^{\beta}\neq1,
\end{align}
and
\begin{align}\label{x6}
e^{\beta}=1+Fe^{\alpha}\alpha'^{2}.
\end{align}
Here, $F$ is an integration constant. After rearranging Eq. \eqref{a7}, we have
\begin{align}\label{x7}
e^{\alpha}=\left[B+C\sqrt{e^{\beta}-1}\right]^{2},
\end{align}
which describes a relationship between the geometric variables $\alpha(r)$ and $\beta(r)$.

Now, we revisit the basic structure of the gravitational decoupling scheme employing the MGD-decoupling. In the previous section, we discuss the effectiveness of gravitational decoupling as a technique to split the seed source into two independent gravitational sectors. This separation is achieved through a mechanism that maintains the essential geometric properties of the underlying problem. Consequently, after decoupling, the respective conservation relations remain valid for each gravitational sector. The covariant conservation of both the gravitational-field sources implies that their interaction is gravitational. However, there stems a question regarding the separation of the gravitational sector fulfilling the original fluid features. For a more profound understanding of this phenomenon, we consider a matter distribution characterized as an isotropic/perfect fluid. This distribution is described by the following SET
\begin{align}\label{s1}
\widetilde{T}_{\mu\nu}=(\widetilde{\rho}+\widetilde{P})\xi_{\mu}\xi_{\nu}-\widetilde{P}g_{\mu\nu},
\end{align}
where $\widetilde{\rho}$ is the energy density and $\widetilde{P}$ is the isotropic pressure. We now extend the gravitational field source $T_{\mu\nu}$ by including the additional field source $\Theta_{\mu\nu}$, decoupled through a dimensionless constant $\chi$ as
\begin{align}\label{s2}
\widetilde{T}_{\mu\nu}\mapsto T_{\mu\nu}=\widetilde{T}_{\mu\nu}+\chi\Theta_{\mu\nu}.
\end{align}
The introduction of this additional field term, $\Theta_{\mu\nu}$, unveils novel physical phenomena that extend beyond the confines of the established theoretical framework of GR. Despite the remarkable success of GR in understanding a vast array of gravitational interactions, including cosmological matters, the solar system, black holes, neutron stars, and even hypothetical structures like wormholes, it is widely acknowledged that modifications or extensions may be necessary to fully incorporate certain gravitational aspects. We now consider a self-gravitating compact object, potentially representing an NS or a quark star, described by the source terms introduced in Eqs. \eqref{a3} and \eqref{s2}. Then, owing to the spherically symmetric metric ansatz and the isotropic nature of the matter distribution, we have $\widetilde{T}^{r}_{r}=\widetilde{T}^{\theta}_{\theta}=\widetilde{T}^{\phi}_{\phi}=\widetilde{P}$. However, if we consider the system \eqref{s2} instead of \eqref{s1}, then the introduction of the new sector may lead to a distinct behavior, wherein $T^{r}_{r}\neq T^{\theta}_{\theta}=T^{\phi}_{\phi}$. Consequently, the relativistic configuration exhibits an anisotropic behavior. The manifestation of this induced anisotropy in the matter content is critically dependent on the particular features of the $\Theta_{\mu\nu}$ field. This field can take the form of a scalar, vector, or tensor field. Furthermore, it could be interpreted as a signature of a dark sector component, such as DM or dark energy \cite{ovalle2018black,cedeno2020gravitational,ovalle2019decoupling}. In order to separate the field sources $\widetilde{T}_{\mu\nu}$ and $\Theta_{\mu\nu}$, it is advantageous to understand how these sources relate to the geometry of spacetime. This correspondence is developed by varying the action
\begin{align}\label{s4}
\mathcal{A}^{\circ}=\mathcal{A}_{\mathrm{EH}}+\mathcal{A}_{\textmd{matter}},
\end{align}
with respect to $g^{\mu\nu}$. In the above expression, $\mathcal{A}_{\mathrm{EH}}$ denotes the usual Einstein-Hilbert action, defined as
\begin{align}\label{s5}
\mathcal{A}_{\mathrm{EH}}=\frac{1}{16\pi}\int d^{4}x\sqrt{|g|}\mathrm{R}.
\end{align}
Here, $\mathrm{R}$ denotes the contraction of the Ricci tensor $R_{\mu\nu}$, commonly referred to as the Ricci scalar, while $g$ represents the trace of $g_{\mu\nu}$.
On the other hand, the matter content is characterized by the following action
\begin{align}\label{s6}
\mathcal{A}_{\textmd{matter}}=\int d^{4}x\sqrt{|g|}\mathscr{L}_{\textmd{matter}},
\end{align}
where $\mathscr{L}_{\textmd{matter}}$ is the matter Lagrangian. Generally, this Lagrangian density could encompass a broader range of fields representing diverse forms of matter distributions, which are not described by GR.
\begin{align}\label{s7}
\mathscr{L}_{\textmd{matter}}=\mathscr{L}_{\widetilde{\textmd{M}}}+\chi\mathscr{L}_{\mathbb{X}},
\end{align}
where $\mathscr{L}_{\widetilde{\textmd{M}}}$ captures the contributions of the seed source, while $\mathscr{L}_{\mathbb{X}}$ encodes the characteristics of the new matter fields. These additional contributions may be regarded as corrections to GR \cite{ovalle2019decoupling}. The gravitational-field equations describing the gravity-matter correspondence are obtained from the variational principle, which reads
\begin{align}\label{s8}
\frac{\delta \mathcal{A}^{\circ}}{\delta g^{\mu\nu}}=0\Rightarrow G_{\mu\nu}\equiv R_{\mu\nu}-
\frac{1}{2}\mathrm{R}g_{\mu\nu}=-8\pi T_{\mu\nu}.
\end{align}
Here, $G_{\mu\nu}$ denotes the usual Einstein tensor and
\begin{align}\label{s9}
T_{\mu\nu}=-2\frac{\delta \mathscr{L}_{\widetilde{\textmd{M}}}}{\delta g^{\mu\nu}}+
g_{\mu\nu}\mathscr{L}_{\widetilde{\textmd{M}}}+\chi\left(-2\frac{\delta \mathscr{L}_{\mathbb{X}}}{\delta g^{\mu\nu}}+
g_{\mu\nu}\mathscr{L}_{\mathbb{X}}\right),
\end{align}
where
\begin{align}\label{s10}
\widetilde{T}_{\mu\nu}=-2\frac{\delta \mathscr{L}_{\widetilde{\textmd{M}}}}{\delta g^{\mu\nu}}+
g_{\mu\nu}\mathscr{L}_{\widetilde{\textmd{M}}},
\end{align}
and
\begin{align}\label{s11}
\Theta_{\mu\nu}=-2\frac{\delta \mathscr{L}_{\widetilde{\textmd{M}}}}{\delta g^{\mu\nu}}+
g_{\mu\nu}\mathscr{L}_{\widetilde{\textmd{M}}}.
\end{align}
The divergence-free nature of $G_{\mu\nu}$ provides the following result
\begin{align}\label{s12}
\nabla_{\mu}G^{\mu\nu}=0\Rightarrow\nabla_{\mu}T^{\mu\nu}=0,
\end{align}
which ensures the covariant conservation of the SET, $T_{\mu\nu}$.
\begin{align}\label{s13}
\nabla_{\mu}T^{\mu\nu}=0\Rightarrow\nabla_{\mu}\widetilde{T}^{\mu\nu}+\chi\nabla_{\mu}\Theta^{\mu\nu}=0.
\end{align}
The solution of the above equation can be described in two ways:
\begin{description}
  \item[(i)] $\nabla_{\mu}T^{\mu\nu}=\nabla_{\mu}\Theta^{\mu\nu}=0$,
  \item[(ii)] $\nabla_{\mu}T^{\mu\nu}=-\nabla_{\mu}\Theta^{\mu\nu}$.
\end{description}
The first option suggests that each source is covariantly conserved, implying a purely gravitational interaction between them. On the other hand, the second option indicates an energy exchange between both the fields \cite{ovalle2022energy,contreras2022energy}. In this investigation, we are mainly concerned with the second case. Consequently, the gravitational system $\{ds^{2},T_{\mu\nu}\}$ produces the following set of differential equations
\begin{align}\label{s14}
G^{t}_{t}&=-8\pi T^{t}_{t}:\frac{1}{r^{2}}-\left(\frac{1}{r^{2}}-\frac{\beta'}{r}\right)e^{\beta}=-8\pi(\widetilde{\rho}
+\chi \Theta^{t}_{t}),
\\\label{s15}
G^{r}_{r}&=-8\pi T^{r}_{r}:-\frac{1}{r^{2}}+\left(\frac{1}{r^{2}}-\frac{\alpha'}{r}\right)e^{\beta}
=-8\pi(-\widetilde{P}+
\chi\Theta^{r}_{r}),
\end{align}
\begin{widetext}
\begin{align}\label{s16}
G^{\theta}_{\theta}&=-8\pi T^{\theta}_{\theta}:\frac{e^{-\beta}}{4}\left(2\frac{\beta'}{r}-2\frac{\alpha'}{r}+\beta'\alpha'
-\alpha'^{2}-2\alpha''\right)=-8\pi(-\widetilde{P}+\chi\Theta^{\theta}_{\theta}),
\end{align}
\end{widetext}
where $G^{\theta}_{\theta}=G^{\phi}_{\phi}$ due to spherical symmetry. The mass function for the self-gravitational, spherically symmetric configuration of matter can be defined by employing the Misner and Sharp scheme as
\begin{align}\label{x1}
R^{3}_{232}=1-e^{-\beta}=\frac{2m}{r},
\end{align}
or, alternatively
\begin{align}\label{x2}
m(r)=4\pi\int^{r}_{0}\left[\rho(x)\right] x^{2}dx,
\end{align}
which can be represented in a particular form accounting for the effects of density homogeneity plus the change induced by density inhomogeneity as
\begin{align}\label{x3}
m=\frac{4\pi}{3}r^{3}\rho-\frac{4\pi}{3}\int^{r}_{0}\left[\rho'(x)\right]x^{3}dx.
\end{align}
Then, by using Eqs. \eqref{s14}-\eqref{x1}, we obtain
\begin{align}\label{x4}
\alpha'=\frac{2m+8\pi r^{3}P_{r}}{(r-2m)}.
\end{align}
The covariant conservation of the gravitational-field source $T^{\mu\nu}$ produces the following result
\begin{align}\label{s17}
0\equiv T^{\mu\nu}_{~~~;\nu}=\frac{dT^{\mu\nu}}{dx^{\nu}}+\Gamma^{\mu}_{\nu\epsilon}T^{\epsilon\nu}
+\Gamma^{\nu}_{\nu\epsilon}T^{\mu\epsilon},
\end{align}
which gives the conservation equation for the spherically symmetric stellar distribution for $\mu=1$, given by
\begin{align}\label{s18}
\frac{dT^{r}_{r}}{dr}&=-\frac{1}{2}g^{tt} \frac{dg_{tt}}{dr}\left(T^{t}_{t}-T^{r}_{r}\right)
-g^{\theta\theta} \frac{d g_{\theta\theta}}{dr}\left(T^{r}_{r}-T^{\theta}_{\theta}\right),
\end{align}
from where do we get
\begin{widetext}
\begin{align}\label{s19}
\widetilde{P}'&=-\frac{\alpha'}{2}(\widetilde{\rho}+\widetilde{P})
+\chi \left[\left(\Theta^{r}_{r}\right)'
-\frac{\alpha'}{2}\left(\Theta^{t}_{t}-\Theta^{r}_{r}\right)
-\frac{2}{r}\left(\Theta^{\theta}_{\theta}-\Theta^{r}_{r}\right)\right].
\end{align}
\end{widetext}
Then, by simple observation, we can write
\begin{align}\label{s20}
&T^{t}_{t}=\widetilde{\rho}+\chi\Theta^{t}_{t}\equiv\rho,
\\\label{s21}
&T^{r}_{r}=-\widetilde{P}-\chi\Theta^{r}_{r}\equiv P_{r},
\\\label{s22}
&T^{\theta}_{\theta}=-\widetilde{P}-\chi\Theta^{\theta}_{\theta}\equiv P_{t},
\end{align}
where clearly we have
\begin{align}\label{b1}
&T^{\mu}_{\nu}=\textmd{Diag}\Big[\rho,-P_{r},-P_{\theta},-P_{\phi}\Big],
\\\label{b2}
&\Theta^{\mu}_{\nu}=\textmd{Diag}\left[\Theta^{t}_{t},-\Theta^{r}_{r},-\Theta^{\theta}_{\theta},
-\Theta^{\phi}_{\phi}\right].
\end{align}
We now employ the GD approach by considering a solution for the seed gravitational-field source $\widetilde{T}_{\mu\nu}$, i.e.,
\begin{align}\label{b3}
T_{\mu\nu}=\widetilde{T}_{\mu\nu}+\chi\cancelto{{0}}{\Theta_{\mu\nu}},
\end{align}
which can be formally written as
\begin{align}\label{b4}
ds^{2}=e^{u(r)}dt^{2}-\frac{dr^{2}}{y(r)}-r^{2}d\Omega^{2}.
\end{align}
The effects emerging from the generic gravitational-field source $\Theta_{\mu\nu}$ can be observed through the geometric deformation of the line element \eqref{b4}, as follows:
\begin{align}\label{b5}
&\alpha(r)\mapsto \alpha(r)=\eta(r)+\chi h(r),
\\\label{s23}
&e^{-\beta(r)}\mapsto e^{-\beta(r)}=y(r)+\chi f(r),
\end{align}
where $h$ and $f$ denote the deformation functions. Henceforth, we exclusively examine the MGD decoupling, characterized by minimal deformation where $h=0$. Consequently, only $e^{-\beta(r)}$ undergoes modification, with $\alpha=\eta$.
Therefore, we summarize the primary characteristics of the MGD-decoupling scheme as follows:
\begin{itemize}
  \item According to \eqref{s14}, one can conclude that the energy density relies solely on $e^{\beta(r)}$ only. Thus, one can conclude that the linear mapping \eqref{s23} appears as the only possibility to decouple the gravitational-field sources $\widetilde{T}_{\mu\nu}$ and $\Theta_{\mu\nu}$. In this context, the linear mapping that introduces deformations only in the radial metric potential is not realizable without considering the extended scenario \cite{ovalle2019decoupling,khan2024complexity}.
  \item The radial metric deformation $f(r)$ should be purely radial for the preservation of spherical symmetry.
  \item The geometric variable $y(r)$ acts as the primary influence on the behavior of $f(r)$. As $y(r)$ is a strictly increasing function, the sign or monotonicity (increasing/decreasing) of $f(r)$ is irrelevant. The crucial aspect is that $y(r)$ must dominate the behavior of $f(r)$ to ensure a physically viable solution. Additionally, $f(r)\rightarrow0$ at the center of the stellar distribution.
  \item Since the metric variable $y(r)$ is closely linked with Misner-Sharp mass $m(r)$, the linear mapping \eqref{s23} extends the classical description of the mass of the self-gravitational compact configuration by introducing an additional component. Thus, the minimally deformed form of $m(r)$ reads
\end{itemize}
\begin{align}\label{b6}
m(r)=\frac{r}{2}\left[1-y(r)-\chi f(r)\right].
\end{align}
Next, employing the linear mapping \eqref{s23} into the gravitational system \eqref{s14}-\eqref{s16} provides the following differential equations for the seed gravitational-field source
\begin{align}\label{x23}
&8\pi\widetilde{\rho}=\frac{1-y}{r^{2}}-\frac{y'}{r},
\\\label{s24}
&8\pi\widetilde{P}=-\frac{1}{r^{2}}+y\left(\frac{1}{r^{2}}+\frac{\alpha'}{r}\right),
\\\label{s25}
&8\pi\widetilde{P}=\frac{y}{4}\left(2\frac{\alpha'}{r}+2\alpha''+\alpha'^{2}\right)
+\frac{y'}{r}\left(\alpha'+\frac{2}{r}\right),
\end{align}
and the corresponding conservation equation reads
\begin{align}\label{s26}
\widetilde{P}'+\frac{\alpha'}{2}(\widetilde{\rho}+\widetilde{P})=0.
\end{align}
Now, the stellar system describing the imprints of additional field source $\Theta_{\mu\nu}$ can be written as
\begin{align}\label{s27}
&8\pi\Theta^{t}_{t}=-\left(\frac{f}{r^{2}}+\frac{f'}{r}\right),
\\\label{s28}
&8\pi\Theta^{r}_{r}=-f\left(\frac{1}{r^{2}}+\frac{\alpha'}{r}\right),
\\\label{s29}
&8\pi\Theta^{\theta}_{\theta}=-\frac{f}{4}\left(2\frac{\alpha'}{r}+2\alpha''+\alpha'^{2}\right)
-\frac{f'}{4}\left(\alpha'+\frac{2}{r}\right),
\end{align}
whose conservation equation likewise reads
\begin{align}\label{s30}
\left(\Theta^{r}_{r}\right)'
-\frac{\alpha'}{2}\left(\Theta^{t}_{t}-\Theta^{r}_{r}\right)
-\frac{2}{r}\left(\Theta^{\theta}_{\theta}-\Theta^{r}_{r}\right)=0.
\end{align}
Here are some key observations regarding the systems representing the seed and $\Theta$ gravitational-field sources:
\begin{enumerate}
  \item The gravitational system encoding the imprints of decoupling fluid $\Theta_{\mu\nu}$ correlates with the quasi-Einstein filed equations; however, they deviate by a factor of $1/r^{2}$.
  \item The independent conservation of both field sectors implies a purely gravitational interaction between them.
      \item The self-gravitational system filled isotropic matter content transitions to an anisotropic state if and only if $\Theta^{r}_{r}\neq\Theta^{\theta}_{\theta}$.
\end{enumerate}
Therefore, the value of the anisotropic scalar for the considered system turns out to be
\begin{align}\label{s31}
\Pi(r,\chi)\equiv P_{r}-P_{t}=\chi(\Theta^{\theta}_{\theta}-\Theta^{r}_{r}),
\end{align}
which clearly describes the role of the $\Theta$ field source in generating anisotropy within the isotropic matter configuration. Importantly, for a physically viable self-gravitational model, one must have $\Pi>0$ throughout the range, $0\leq r\leq R$.

The seed system is characterized by Eqs. \eqref{x23}-\eqref{s25} is already established by the seed metric. However, the $\Theta$-system is not closed as described by the set of Eqs. \eqref{s27}-\eqref{s29}. The inclusion of $\Theta$-field source introduces four unknowns: $\{\Theta^{t}_{t},\Theta^{r}_{r},\Theta^{\theta}_{\theta},f(r)\}$. However, only three equations govern their behavior. Consequently, it becomes necessary to impose an additional constraint to establish a well-defined system. When it comes to the closure of the $\ Theta$ system, various approaches have been implemented, leading demonstrably to well-behaved interior solutions in most cases. Among different gravitational techniques, the typical ones include: i) the mimic constraint methodology \cite{ovalle2018anisotropic,gabbanelli2018gravitational,morales2018charged,estrada2019gravitational}; ii) enforcing a suitable deformation function $f(r)$ \cite{maurya2019generalized,morales2018compact}; and iii) employing a regularity requirement on the altered segment driven by the renowned anisotropy condition as elaborated in \cite{abellan2020regularity}.

However, we consider an alternative scheme to close the $\Theta$-gravitational field source.
As highlighted in \cite{ovalle2019decoupling} and further investigated in the context of cosmology \cite{cedeno2020gravitational}, the enigmatic birth of the $\Theta$-gravitational sector may be theoretically grounded by postulating that this field constitutes one of the mysterious, unseen constituents of our universe. By drawing a parallel to DM, we consider a hypothesis suggesting a potential connection between the $\Theta$-field and a DM constituent. Building on this conjecture, we associate the temporal aspect of the $\Theta$-field with a DM density distribution. In this direction, we have explicitly assumed $\Theta^{t}_{t}=\rho$, as elaborated upon in the subsequent section.

Now, the definition of mass function for the seed gravitational sector $\widetilde{T}_{\mu\nu}$ reads
\begin{align}\label{y1}
m_{s}(r)=\frac{r}{2}(1-y)=4\pi\int^{r}_{0}\left[\widetilde{\rho}(x)\right]x^{2}dx,
\end{align}
which implies
\begin{align}\label{y2}
m=m_{s}-\chi\frac{r}{2}f(r)\equiv m_{s}+\chi m_{\Theta},
\end{align}
where the mass functions $m_{s}$ and $m_{\Theta}$ can be expressed as
\begin{align}\label{y3}
&m_{s}=\frac{4\pi}{3}\widetilde{\rho} r^{3}-\frac{4\pi}{3}\int^{r}_{0}[\widetilde{\rho}(x)]x^{3}dx,
\\\label{y4}
&m_{\Theta}=\frac{4\pi}{3}\Theta^{t}_{t}r^{3}-\frac{4\pi}{3}\int^{r}_{0}[\Theta^{t}_{t}(x)]x^{3}dx.
\end{align}
The generic physical and mathematical features associated with the minimally deformed Finch-Skea metric will be analyzed in the following sections.

\section{Deformed Schwarzschild Vacuum and Matching Conditions}

Let us consider the matching constraints that guarantee the smooth matching between the interior $(0\leq r \leq R)$ of a stellar configuration and the outer geometry $(r>R)$ at the boundary surface $\Sigma:r=R$. We describe the interior geometry using the generic metric \eqref{a3}. This metric can be expressed in terms of the MGD transformation \eqref{s23} as follows
\begin{align}\label{i1}
ds^{2}=e^{\alpha^{-}}dt^{2}-\left(1-\frac{2m}{r}\right)^{-1} dr^{2}-r^{2}d\Omega^{2},
\end{align}
with $m=m_{s}-\frac{r}{2}\chi f$. However, it is important to note that the exterior metric may deviate from a strict vacuum state. In general, the inclusion of the additional field source, $\Theta_{\mu\nu}$, may produces new fields that do not exist in a standard vacuum. Therefore, the generic form of the outer metric can be defined as
\begin{align}\label{i2}
ds^{2}=e^{\alpha^{+}}dt^{2}-e^{\beta^{+}}dr^{2}-r^{2}d\Omega^{2},
\end{align}
where metric functions $e^{\alpha^{+}}$ and $e^{\beta^{+}}$ can be determined by solving the effective four-dimensional exterior Einstein stellar structure equations
\begin{align}\label{i3}
R_{\mu\nu}-\frac{1}{2}R^{\mu}_{\mu}\textsl{g}_{\mu\nu}=\chi\Theta_{\mu\nu}.
\end{align}
Applying the MGD decoupling technique reduces the aforementioned stellar structure equations to the system of Eqs. \eqref{s27}--\eqref{s29}, where the metric variable $\alpha$ is determined by the Schwarzschild solution. Now, the continuity of the first fundamental form at the stellar surface reads
\begin{align}\label{i4}
\left[ds^{2}\right]_{\Sigma}=0,
\end{align}
where $\left[\mathcal{G}\right]\equiv \mathcal{G}(r\rightarrow R^{+})-\mathcal{G}(r\rightarrow R^{-})
\equiv \mathcal{G}_{R}^{+}-\mathcal{G}_{R}^{-}$, for any function $\mathcal{G}=\mathcal{G}(r)$, which gives
\begin{align}\label{i5}
\alpha^{-}(R)=\alpha^{+}(R),
\end{align}
and
\begin{align}\label{i6}
1-\frac{2M_{\textmd{Sch}}}{R}+\chi f_{R}=e^{-\beta^{+}(R)},
\end{align}
where $M_{\textmd{Sch}}=m_{s}(R)$ and $f_{R}$ represents the MGD deformation at the stellar surface.
Similarly, the continuity of the second fundamental form is expressed by
\begin{align}\label{i7}
\left[G_{\mu\nu} r^{\nu}\right]_{\Sigma}=0,
\end{align}
where $r_{\mu}$ denotes the unit-radial radial. Then, using the above expression in gravitational equations of motion \eqref{s8}, we have
\begin{align}\label{i8}
\left[T_{\mu\nu} r^{\nu}\right]_{\Sigma}=0,
\end{align}
which implies
\begin{align}\label{i9}
\left[\widetilde{P}-\chi\Theta^{r}_{r}\right]_{\Sigma}=0,
\end{align}
\begin{align}\label{i10}
\widetilde{P}_{R}-\chi\left(\Theta^{r}_{r}\right)_{R}^{-}=-\chi\left(\Theta^{r}_{r}\right)_{R}^{+}.
\end{align}
This expression signifies an important result describing the second fundamental form corresponding to the Einstein stellar structure Eqs. \eqref{s8} and the Schwarzschild vacuum filled with the decoupling fluid $\Theta^{+}_{\mu\nu}$, namely, $\{\Theta^{+}_{\mu\nu}\neq0,T^{+}_{\mu\nu}=0\}$. Then, substituting Eq. \eqref{s45} for the interior geometry in Eq. \eqref{i10}, we get
\begin{align}\label{i11}
\widetilde{P}_{R}+\frac{\chi f_{R}}{8\pi}\left(\frac{1}{R^{2}}+\frac{\alpha'_{R}}{R}\right)
=-\chi\left(\Theta^{r}_{r}\right)_{R}^{+},
\end{align}
where $\alpha'_{R}=\left(\frac{\partial\alpha^{-}}{\partial r}\right)_{r=R}$. Now, substituting Eq. \eqref{s45} for the exterior geometry in Eq. \eqref{i11}, we have
\begin{align}\label{i12}
\widetilde{P}_{R}+\frac{\chi f_{R}}{8\pi}\left(\frac{1}{R^{2}}+\frac{\alpha'_{R}}{R}\right)
=\frac{\chi s^{\ast}_{R}}{8\pi}\left[\frac{1}{R^{2}}+\frac{2M^{\ast}}{R^{3}}\left(1-\frac{2M^{\ast}}{R}\right)^{-1}\right].
\end{align}
Here, the function $s^{\ast}_{R}$ captures the geometric deformation associated with the outer Schwarzschild metric due to the addition of the extra field source $\Theta_{\mu\nu}$, that is,
\begin{align}\label{i13}
ds^{2}=\left(1-\frac{2M^{\ast}}{R}\right)dt^{2}-\left(1-\frac{2M^{\ast}}{R}+\chi s^{\ast}\right)^{-1}dr^{2}-
r^{2}d\Omega^{2}.
\end{align}
Finally, Eqs. \eqref{i5}, \eqref{i6} and \eqref{i12} describe the necessary and sufficient conditions that must be satisfied to ensure a smooth matching between the minimally deformed inner geometry \eqref{i1} and the outer geometry describing a spherically symmetric vacuum field, with the minimally deformed Schwarzschild metric \eqref{i13}. It is crucial to understand that the exterior metric may not be entirely free of matter or energy. The source term, $\Theta_{\mu\nu}$, permits the presence of various fields within this region. The junction constraint \eqref{i13} leads to an important result: if the outer geometry is described by the exact Schwarzschild metric, then $s^{\ast}(r)=0$, which in turn provides
\begin{align}\label{i15}
P_{R}=\widetilde{P}_{R}+\frac{\chi f_{R}}{8\pi}\left(\frac{1}{R^{2}}+\frac{\alpha'_{R}}{R}\right)=0.
\end{align}
Thus, the essential condition for a stellar configuration to be in equilibrium within a true Schwarzschild vacuum is that the effective pressure at its surface must be zero. An interesting result emerges when the interior geometric deformation ($f_R<R$) weakens the gravitational field and is positive. In this scenario, a standard outer Schwarzschild vacuum can only be compatible with a non-zero inner source term $\Theta_{\mu\nu}$ if the pressure of the isotropic stellar matter at the star's surface is negative ($\widetilde{P}_{R}<0$). This implies the presence of regular matter with a solid crust, as discussed in reference \cite{ovalle2015brane}. Alternatively, we can achieve the pressure condition $\widetilde{P}_{R}=0$ through a different approach. Instead of directly solving for it, we can impose a constraint on Eq. \eqref{i15}, that is, $\beta\left(\Theta^{r}_{r}\right)^{-}_{R}\thicksim\widetilde{P}_{R}$ in Eq. \eqref{i10}. Consequently, the interior deformation, $f_{R}$, vanishes.

\section{Einasto Density Inspired Minimally Deformed Self-gravitational Model}

This section captures the formulation of a minimally deformed DM model for self-gravitational distributions incorporating Einasto density parametrization. We begin by considering the seed metric employed throughout this work. Subsequently, we construct the self-gravitational structures characterizing the seed and $\Theta$-field sources with the help of the Einasto DM profile.
Building upon the groundwork established previously, we extend our understanding of time-independent, isotropic stellar spheres, which satisfy GR-field equations, towards anisotropic domains. This extension requires incorporating elements beyond GR. For this purpose, we have considered the Adler-Finch-Skea ansatz as a seed metric. This model has been employed to represent various self-gravitational compact configurations, including neutron stars and quark stars, by incorporating an isotropic matter content. In this direction, Tello-Ortiz \cite{tello2020class} \emph{et al.} have constructed relativistic anisotropic stellar systems by employing the combination of MGD decoupling with the Class I approach. This is achieved by considering the Adler-Finch-Skea solution as a seed metric. Furthermore, Maurya \emph{et al.} \cite{maurya2020egd} explored realistic anisotropic fluid sphere-like stellar structures within the context of the extended MGD approach by considering the Class I method. To assess the feasibility of this approach, they used the hybrid Adler–Finch–Skea solution as the seed space-time solution. Then, we have the following Class I seed metric
\begin{align}\label{s37}
\alpha(r)=\ln\left[\left(A+Br^{2}\right)^{2}\right],
\end{align}
\begin{align}\label{s38}
y(r)=\frac{1}{1+Cr^{2}}.
\end{align}
Then, the fluid configuration filling the stellar interior, characterized by the above-mentioned geometry, reads
\begin{align}\label{s39}
&8\pi\widetilde{\rho}(r)=\frac{C(3+Cr^{2})}{(1+Cr^{2})^{2}},
\\\label{n40}
&8\pi \widetilde{P}(r)=\frac{4B-C(A+Br^{2})}{(A+Br^{2})(1+Cr^{2})}.
\end{align}
Notably, this isotropic stellar model corresponds to the GR-relativistic equations of motion in the absence of the decoupling fluid $\Theta_{\mu\nu}$. We determine the constants $A$, $B$, and $C$ appearing in Eqs. \eqref{s37}--\eqref{n40} by applying the junction constraints defined in Eqs. \eqref{i4} and \eqref{i7}, between the inner and outer geometry. According to these constraints, the values of $A$, $B$, and $C$ termed as
\begin{align}\label{i16}
\frac{A}{BR^{2}}=&\frac{2R-3M_{\textmd{Sch}}}{M_{\textmd{Sch}}},
\\\label{i17}
B^{2}R^{4}=&\frac{M^{2}_{\textmd{Sch}}}{4R(R-2M_{\textmd{Sch}})},
\\\label{i18}
\frac{2(1+CR)}{CR}=&\frac{R}{M_{\textmd{Sch}}}.
\end{align}
Notably, Eq. \eqref{i18} guarantees a smooth geometric transition at $r=R$. However, this expression will change when we include the additional source term, $\Theta_{\mu\nu}$. Let us now turn on the deformation parameter $\chi$ within the interior stellar configuration. The radial and metric potentials (defined in Eqs. \eqref{s37} and \eqref{s23}, respectively), whereas the additional field source $\Theta_{\mu\nu}$ and the deformation parameter $f(r)$ are interconnected through the stellar structure Eqs. \eqref{s27}--\eqref{s29}. Therefore, additional information must be provided to close the system of Eqs. \eqref{s27}--\eqref{s29}. We have multiple strategies to fully solve this system. One possibility is to use an EoS to define the relationship between density and pressure within the source term $\Theta_{\mu\nu}$. An alternative method is to impose a restriction based on physical principles on the interior deformation $f(r)$. Importantly, we must ensure that any solution we find remains physically realistic. This is a significant challenge, and the following section addresses it by presenting three new, exact solutions that are demonstrably compatible with physical principles.

\subsection{Mimic Constraint for Density}

One potential approach that leads to physically stellar models is the mimic constraint for the density.
This constraint is defined as
\begin{align}\label{s40}
\rho(r)=\Theta^{t}_{t}(r),
\end{align}
which gives rise to the following equation
\begin{align}\label{s41}
\frac{df}{dr}+\frac{f}{r}=-\rho_{0}re^{-\left(\frac{r}{h}\right)^{1/n}},
\end{align}
whose solution provides
\begin{align}\label{s42}
f(r)=\frac{n\rho_{0}}{r}h^{3}\Gamma\left[3n,\left(\frac{r}{h}\right)^{1/n}\right]
+\frac{D_{0}}{r},
\end{align}
where $\Gamma(p,x)=\int^{\infty}_{x}t^{p-1}e^{-t}dt$ symbolize the incomplete Gamma function. Furthermore, $D_{0}$ is an integration constant. To ensure a solution that remains finite at the origin and satisfies $f(0)=0$, we set $D_{0}=0$. Then, the modified form of the radial metric potential becomes
\begin{align}\label{s43}
e^{-\beta(r)}=y(r)+\frac{\chi\rho_{0}}{r}nh^{3}\Gamma\left[3n,\left(\frac{r}{h}
\right)^{1/n}\right].
\end{align}
Then, the continuity of the first form leads to the following results
\begin{align}\label{s44}
(A+BR^{2})^{2}=1-\frac{2M^{\ast}}{R},
\\\label{s45}
\mu(R)+\frac{\chi\rho_{0}}{r}nh^{3}\Gamma\left[3n,\left(\frac{r}{h}
\right)^{1/n}\right]=1-\frac{2M^{\ast}}{R}.
\end{align}
On the other hand, the continuity of the second fundamental form provides
\begin{align}\label{s46}
P_{\mathrm{R}}-\beta(\Theta^{r}_{r})_{R}^{-}=0,
\end{align}
which implies
\begin{widetext}
\begin{align}\label{s47}
C=\frac{4BR^{3}+h^{3}n\chi \rho_{0}(A+5BR^{2})\Gamma\left[3n,(\frac{R}{h})^{\frac{1}{n}}\right]}
{R^{3}(A^{2}+B^{2}R^{2})-h^{3}n\chi \rho_{0}R^{2}(A+5BR^{2})\Gamma\left[3n,(\frac{R}{h})^{\frac{1}{n}}\right]}.
\end{align}
\end{widetext}
Next, by using the expression \eqref{s45}, we get
\begin{align}\label{s48}
\frac{2M^{\ast}}{R}=\frac{2M_{\textmd{Sch}}}{R}-\frac{\chi\rho_{0}}{r}nh^{3}\Gamma\left[3n,\left(\frac{r}{h}
\right)^{1/n}\right].
\end{align}
Now, the combination of the expressions \eqref{y1} and \eqref{s48} produces
\begin{widetext}
\begin{align}\label{s49}
(A+BR^{2})^{2}=1-\frac{2M_{\textmd{Sch}}}{R}+\frac{\chi\rho_{0}}{r}nh^{3}\Gamma\left[3n,\left(\frac{r}{h}
\right)^{1/n}\right].
\end{align}
\end{widetext}
We identify Eqs. \eqref{s47}--\eqref{s49} as the necessary and sufficient conditions for the continuity between the interior and the exterior metric. Finally by employing Einstein Eq. \eqref{s40} along with the Eqs. \eqref{s21} and \eqref{s47}, we find
\begin{widetext}
\begin{align}\label{s50}
P_{r}(r,\chi)=&\frac{h^{3}n\chi\rho_{0}}{r^{3}(A+Br^{2})(A+BR^{2})R^{3}}\left[-R^{3}\left(A^{2}+5r^{2}B^{2}R^{2}
+AB(5r^{2}+R^{2})\right)\Gamma\left[3n,\left(\frac{r}{h}
\right)^{1/n}\right]\right.
\\\label{s51}
-&\left.r^{3}\left(A^{2}+5r^{2}B^{2}R^{2}+AB(r^{2}+5R^{2})\right)\Gamma\left[3n,\left(\frac{R}{h}
\right)^{1/n}\right]\right].
\end{align}
\end{widetext}
\begin{align}\label{b51}
\rho(r,\chi)=(1+\chi)\rho(r),
\end{align}
on the effective tangential pressure becomes
\begin{widetext}
\begin{align}\label{s52}
P_{t}(r,\chi)&=P_{r}(r)+\left[(A+3Br^{2})e^{-(\frac{r}{h})^{\frac{1}{n}}}+nh^{3}R^{3}
(A-5Br^{2})\Gamma\left[3n,\left(\frac{r}{h}
\right)^{\frac{1}{n}}\right]-2nr^{3}h^{3}(A+5Br^{2})\Gamma\left[3n,\left(\frac{R}{h}
\right)^{\frac{1}{n}}\right]\right]
\\\label{253}
&\times\frac{\chi\rho_{0}}{2}/[{R^{3}h^{3}(A+Br^{2})}],
\end{align}
\end{widetext}
whereas the anisotropic factor reads
\begin{widetext}
\begin{align}\label{s53}
\Pi(r,\chi)=\frac{\chi\rho_{0}}{2}\left[\frac{(A+3Br^{2})e^{-(\frac{r}{h})^{1/n}}+nh^{3}R^{3}
(A-5Br^{2})\Gamma\left[3n,\left(\frac{r}{h}
\right)^{1/n}\right]-2nr^{3}h^{3}(A+5Br^{2})\Gamma\left[3n,\left(\frac{R}{h}
\right)^{1/n}\right]}{R^{3}h^{3}(A+Br^{2})}\right].
\end{align}
\end{widetext}

\section{\label{sec:level4}Physical Analysis}

\begin{figure*}
\includegraphics{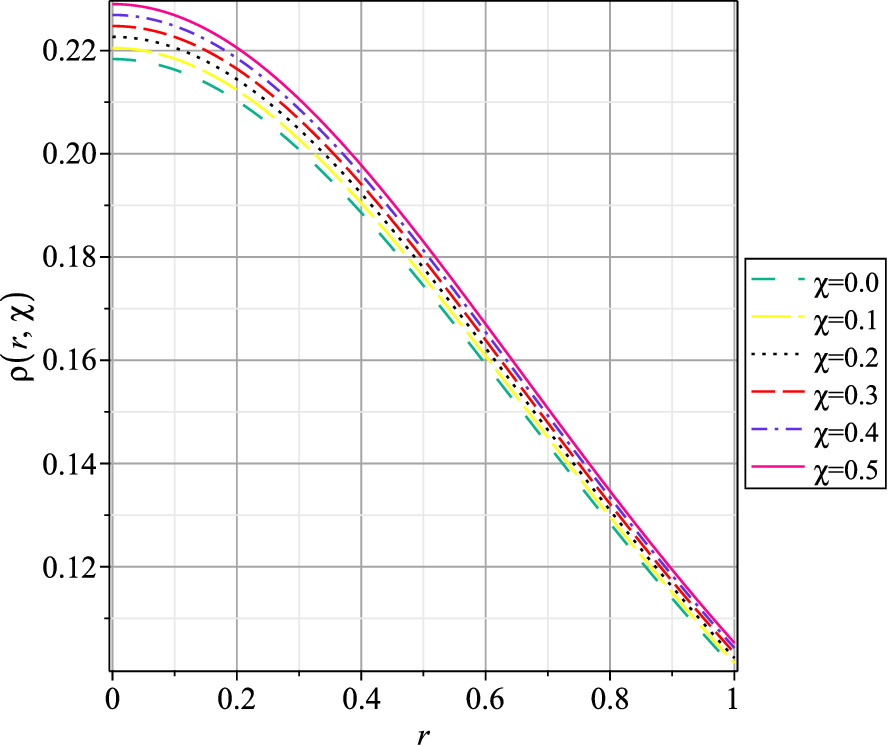}
\caption{\label{1f}Behavior of $\rho$ for the minimally deformed Adler-Finch-Skea model against radial distance $r$ for various $\chi$ values.}
\end{figure*}

\begin{figure*}
\includegraphics{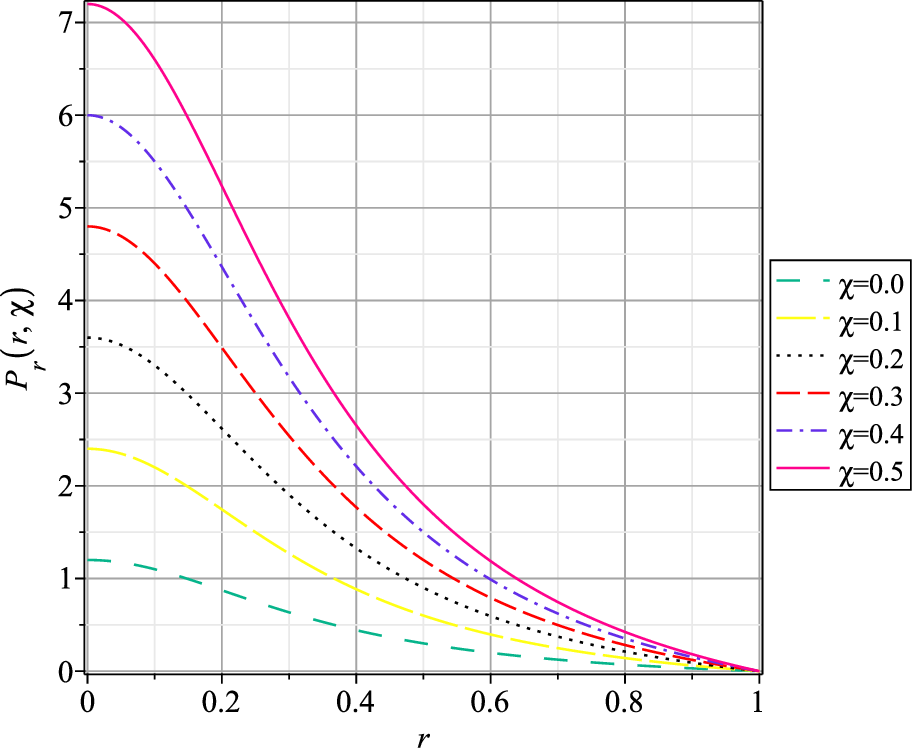}
\caption{\label{2f}Behavior of $P_{r}$ subject to the Adler-Finch-Skea model against radial distance $r$ for various $\chi$ values.}
\end{figure*}

\begin{figure*}
\includegraphics{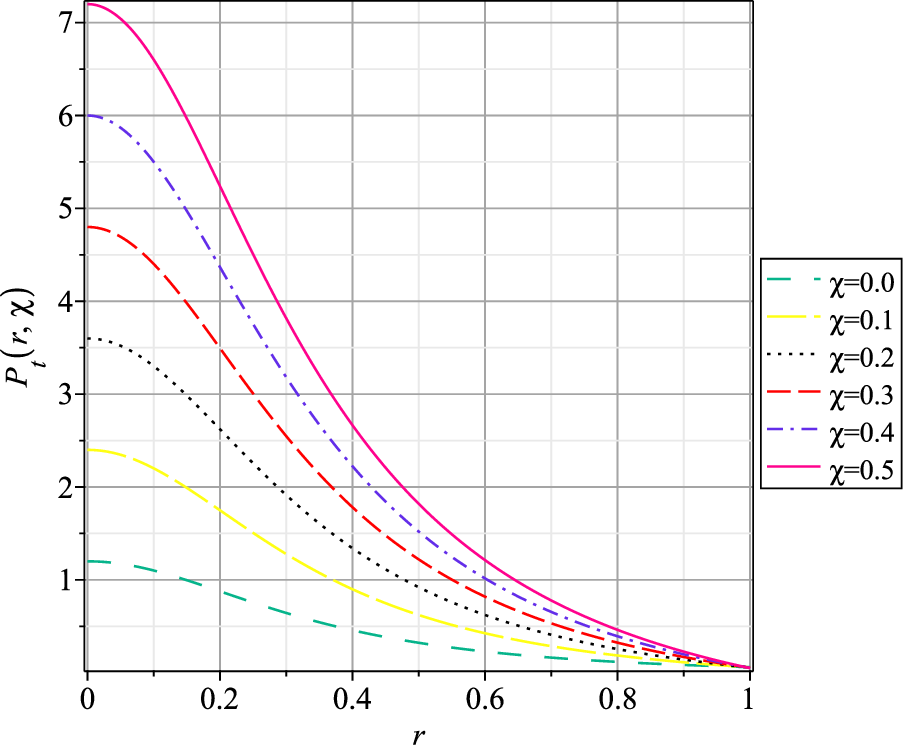}
\caption{\label{3f}Behavior of $P_{t}$ subject to the Adler-Finch-Skea model against radial distance $r$ for various $\chi$ values.}
\end{figure*}

\begin{figure*}
\includegraphics{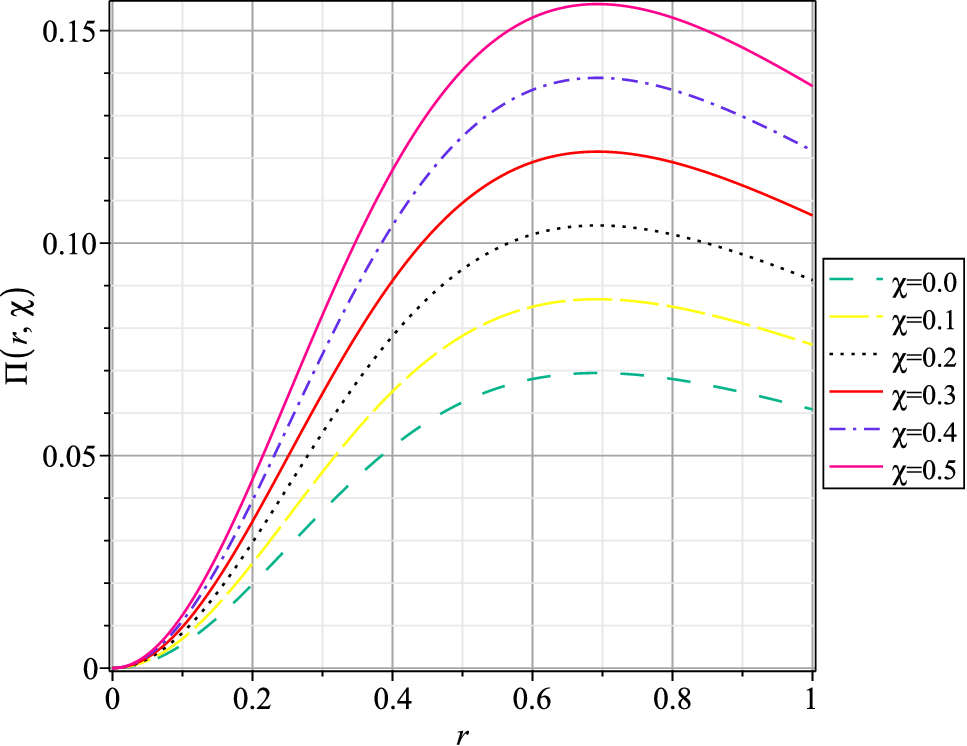}
\caption{\label{4f}Behavior of $\Pi$ subject to the Adler-Finch-Skea model against radial distance $r$ for various $\chi$ values.}
\end{figure*}

 We show the physical behavior of energy density ($\rho$), radial pressure ($P_{r}$), tangential pressure ($P_{t}$), and anisotropic factor ($\Pi$) of an anisotropic fluid sphere as a function of radial distance $r$ in FIGs. \ref{1f}-\ref{4f}, respectively. This analysis is performed for various values of the decoupling constant $\chi$. The profiles of the structural variables $\{\rho, P_{r}, P_{t}, \Pi\}$ satisfy the following essential criteria for a physically valid stellar model:
\begin{enumerate}
  \item Positivity and Finiteness: All three matter variables $\{\rho, P_{r}, P_{t}\}$ remain strictly positive and finite throughout the stellar interior ($0\leq r \leq R$).
  \item Central Peak and Monotonic Decrease: All three matter variables exhibit a maximum value at the center and subsequently exhibit a monotonic decrease as $r$ increases.
\end{enumerate}
From FIG. \ref{2f}, it is clear that the condition for $P_{r}$ to
vanish at the star's surface ($r=R$) is fulfilled. Additionally,
FIG. \ref{4f} illustrates that the $\Pi>0$ increases with
an increase in $r$. This observation is crucial as the positive
behavior of $\Pi$ contributes to enhancing the stability of the
self-gravitational compact structure by preserving hydrostatic
equilibrium. We will now examine the effect of the decoupling
parameter $\chi$ on the structural variables $\{\rho, P_{r}, P_{t},
\Pi\}$ as displayed in FIGs. \ref{1f}-\ref{4f}. It is observed that
the $\rho$ consistently shifts towards higher values throughout the
self-gravitational dark object for each incremental change in
$\chi$. Consequently, the inclusion of the new field source
$\Theta_{\mu\nu}$, representing DM in our model, leads to the
formation of highly dense self-gravitational structures. The
behavior of $P_{r}$ exhibits two distinct patterns, as illustrated
in FIG. \ref{2f} with respect to increasing values of $\chi$. The
value of $P_{r}$ increases in the central region of the stellar
region for the given values of $\chi$. Similar behavior is observed
for $P_{t}$, as shown in FIG. \ref{3f}. However, $P_{t}$ converges
to finite non-zero values at the stellar surface for each value of
$\chi$. This observation is significant for the anisotropic stellar
distribution, where the pressure in both the radial and transverse
directions remains unaffected by $\chi$ values, mimicking the
behavior observed in isotropic systems. FIG. \ref{4f} indicates that
the presence of anisotropy introduced by DM increases near the
surface and converges near the stellar's center as $\chi$ increases.
Consequently, the influence of $\chi$ on the star's anisotropy
steadily rises from the center to the surface of the astrophysical
object.

\begin{figure*}
\includegraphics{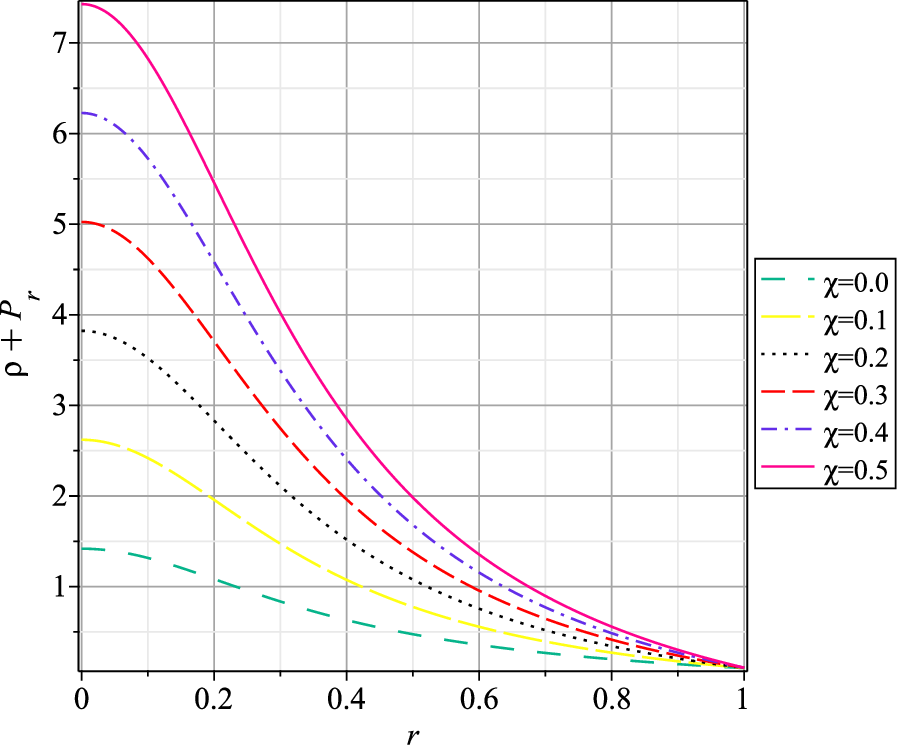}
\caption{\label{5f}Behavior of WEC, $\rho+P_{r}$, versus radial variable $r$ for various $\chi$ values.}
\end{figure*}
\begin{figure*}[t]
\includegraphics{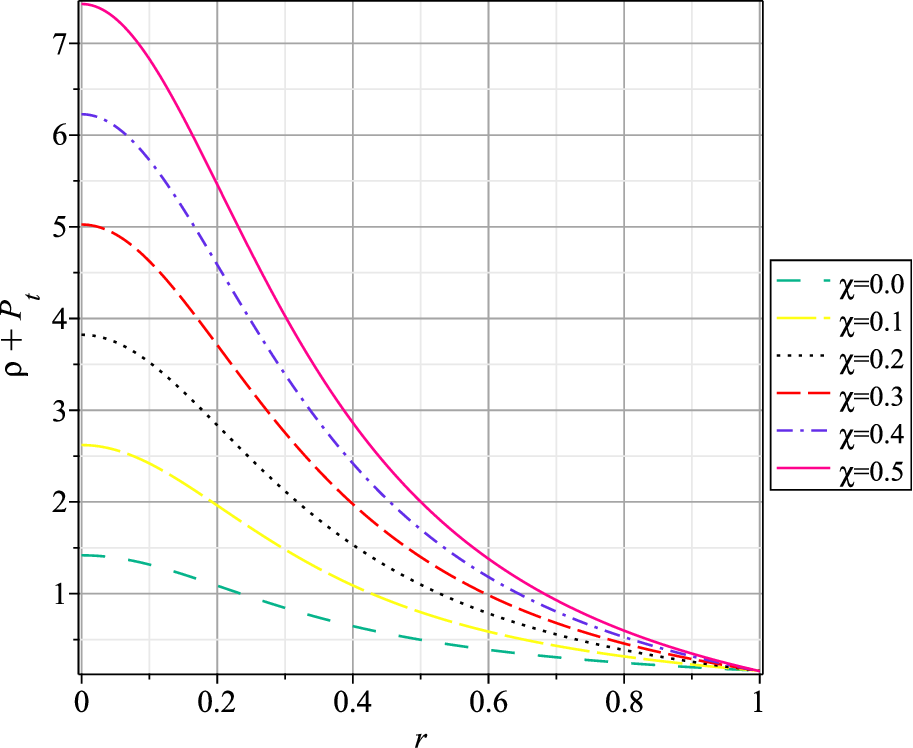}
\caption{\label{6f}Behavior of WEC, $\rho+P_{t}$, versus radial variable $r$ for various $\chi$ values.}
\end{figure*}
\begin{figure*}[t]
\includegraphics{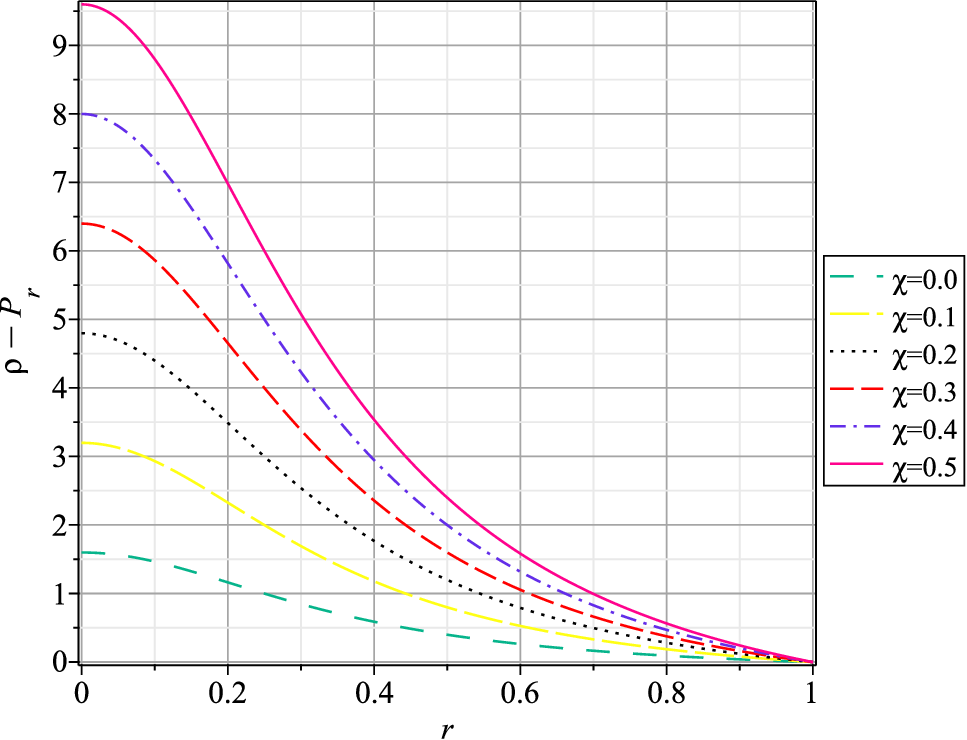}
\caption{\label{7f}Behavior of DEC, $\rho-P_{r}$ versus radial variable $r$ for various $\chi$ values.}
\end{figure*}
\begin{figure*}[t]
\includegraphics{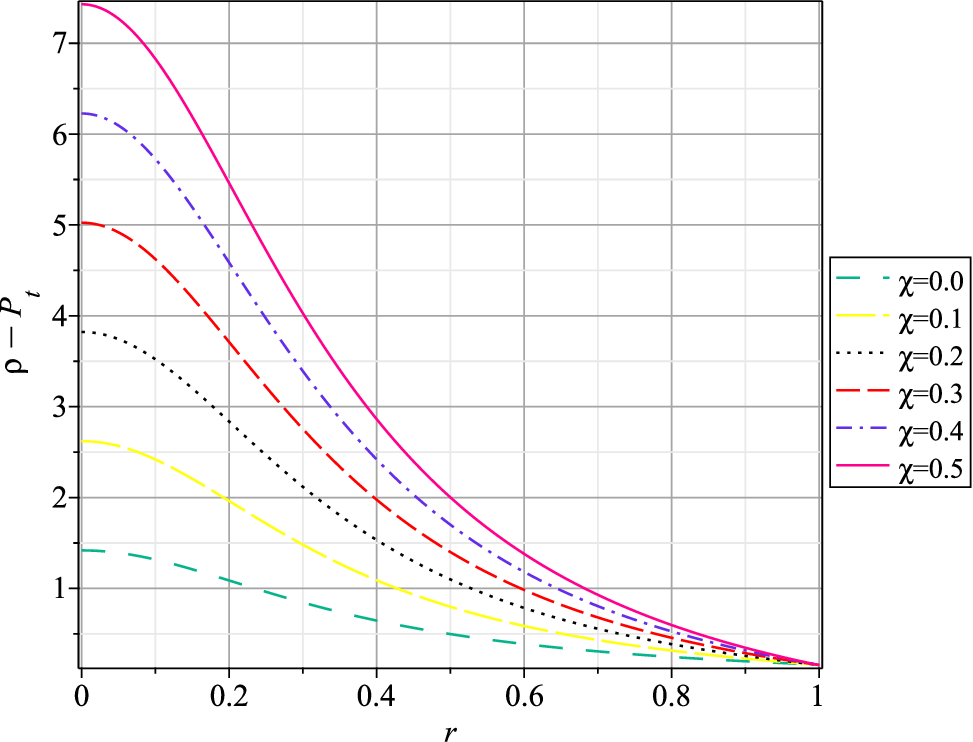}
\caption{\label{8f}Behavior of DEC, $\rho-P_{t}$, versus radial variable $r$ for various $\chi$ values.}
\end{figure*}
\begin{figure*}[t]
\includegraphics{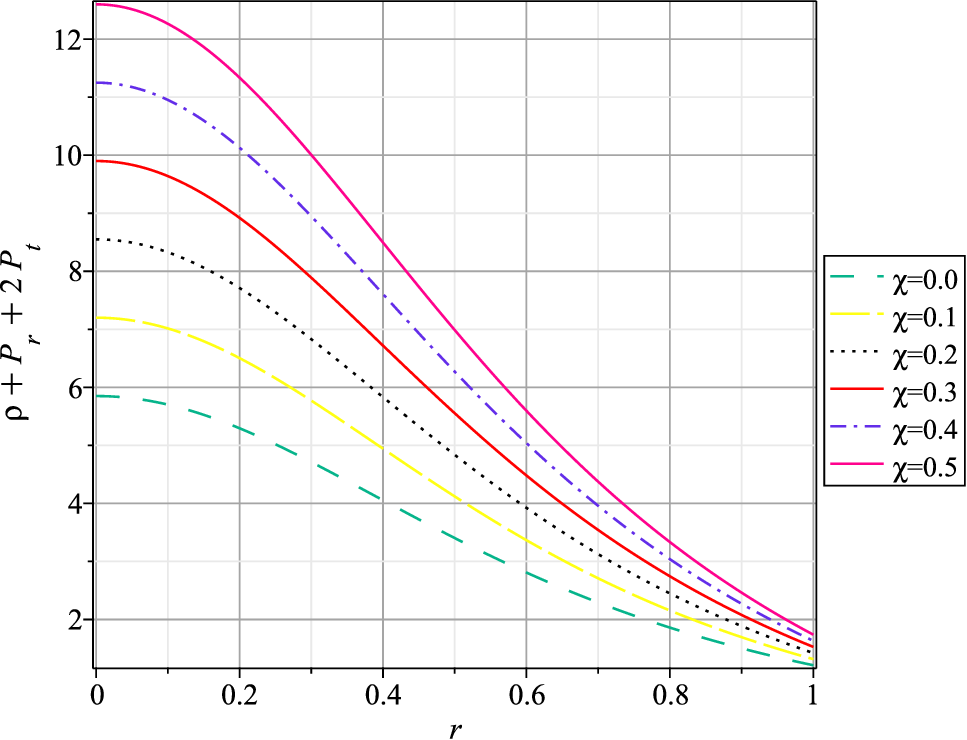}
\caption{\label{9f}Behavior of SEC, $\rho+P_{r}+2P_{t}$, against the radial variable $r$ for various $\chi$ values.}
\end{figure*}
\begin{figure*}[t]
\includegraphics{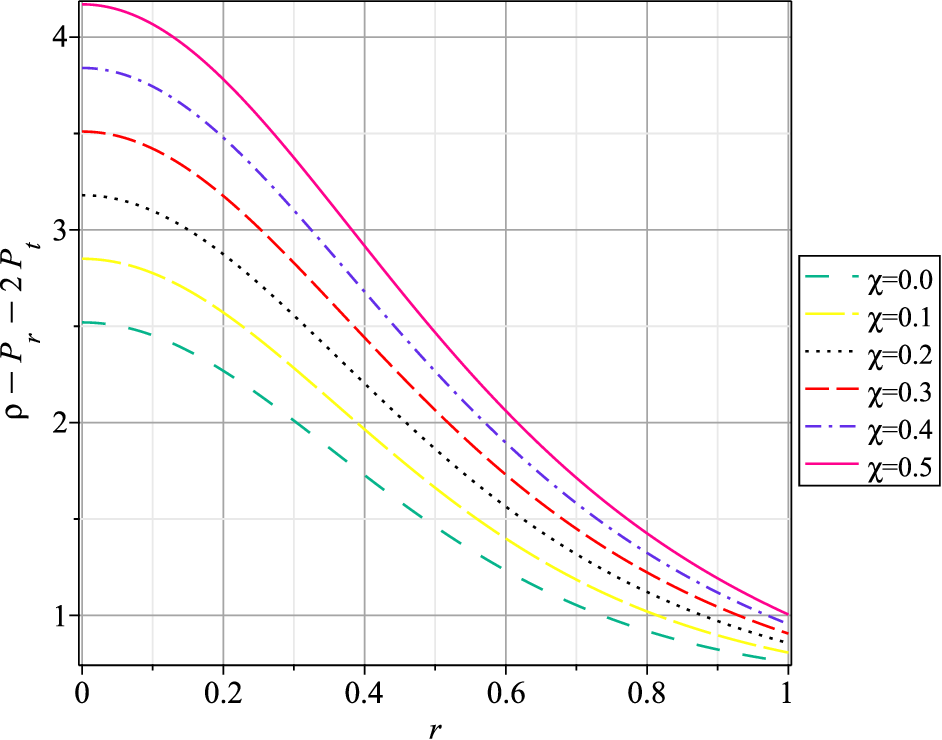}
\caption{\label{10f}Behavior of TEC, $\rho-P_{r}-2P_{t}$, versus $r$ for different $\chi$ values.}
\end{figure*}

{\subsection{Stability of Stellar Distribution through Adiabatic Index}}

{One useful tool to measure the stability of relativistic stellar configurations is the adiabatic index ($\Gamma$). It affects several aspects of structure, stability, and stellar evolution, making it an essential component of these relativistic distributions. A system obeys a stiff equation of state (EoS) when a significant increase in pressure is closely correlated with an increase in the energy density of the system. In comparison to a configuration with a soft equation of state (EoS), a structure associated with a stiff EoS is more stable and more difficult to compress. The parameter $\Gamma$ is used to quantify the stiffness of the EoS. Mathematically, it is represented by
\begin{align}\label{w1}
\Gamma=\frac{\rho+P_{r}}{P_{r}}\frac{dP_{r}}{d\rho},
\end{align}
According to the stability criterion established by Heintzmann and Hillebrandt \cite{heintzmann1975neutron}, a relativistic stellar model is considered stable if its adiabatic index is greater than $4/3$. The behavior of $\Gamma$ is displayed in FIG. \ref{w1}, which shows that anisotropic distribution returns to its equilibrium state after minor density fluctuations corresponding to the different values of $\chi$. The maximum value of $\Gamma$ is observed to occur at the stellar surface, where the radial pressure is zero.}
\begin{figure*}[t]
\includegraphics{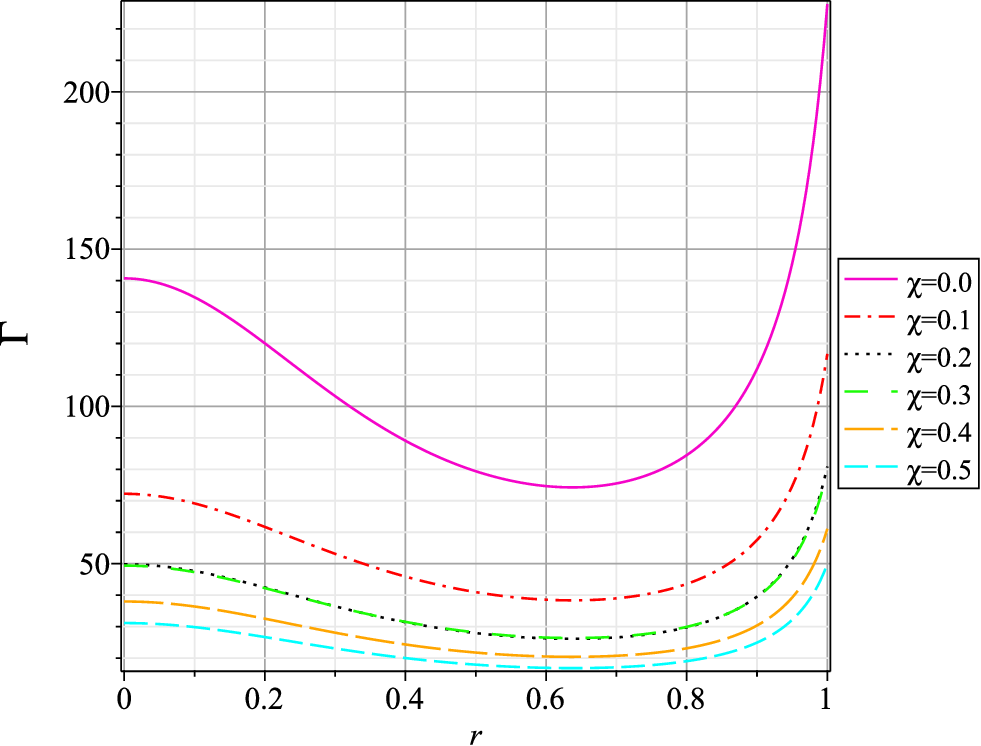}
\caption{\label{w1}Behavior of adiabatic index versus radial variable $r$ for different $\chi$ values.}
\end{figure*}
{\subsection{The Mass-radius Ratio}}

{The effective role of MGD in translating matter configurations beyond the trivial nature, such as a perfect fluid, is described in Section \textbf{II}. Moreover, this scheme alters the stellar geometry and mass distribution while preserving the original symmetry of the seed metric.
The mass function for the self-gravitational, spherically symmetric configuration of matter can be defined by employing the Misner and Sharp scheme as
\begin{align}\label{w2}
m(r)=4\pi\int^{r}_{0}\left[\rho(x)\right] x^{2}dx=4\pi\int^{r}_{0}\left[\widetilde{\rho}(x)+\Theta^{t}_{t}\right] x^{2}dx,
\end{align}
or, alternatively
\begin{align}\label{w3}
R^{3}_{232}=1-e^{-\beta}=\frac{2m}{r},
\end{align}
which gives
\begin{align}\label{w3}
m(r)=\frac{r}{2}\left[1-e^{-\beta}\right]=\frac{r}{2}\left[1-y(r)-\chi f(r)\right].
\end{align}
Subsequently, the total mass is evaluated at $r = R$ and expressed as
\begin{align}\label{w4}
M=m(R)=\frac{R}{2}[1-y(R)]-\chi\frac{R}{2}f(R),
\end{align}
which can be represented as
\begin{align}\label{w5}
M=M_{\mathrm{GR}}-\chi\frac{R}{2}f(R),
\end{align}
where $M_{\mathrm{GR}}$ denotes  the values of total relativistic mass associated with the stellar system within GR. Then, Eq. \eqref{w5} provides the following expression for the mass-radius ratio (compactness factor) $\frac{M}{R}$:
\begin{align}\label{w6}
2\frac{M}{R}=2\frac{M_{\mathrm{GR}}}{R}-\chi f(R).
\end{align}
Several important observations can be made at this point.
\begin{itemize}
  \item The analysis of Eq. \ref{w5} shows that the mass function associated with the seed source increases if the second term is positive. The same applies to the mass-radius ratio $\frac{M}{R}$ in Eq. \eqref{w6}. Therefore, within the context of MGD-decoupling, an additional packing of mass is feasible. This increased packing of mass in self-gravitational distributions has previously been investigated by considering the effect of the Kalb-Ramond field on the matter field \cite{chakraborty2018packing}. The primary distinction between this method and the current one, however, is that the exterior metric is impacted by the Kalb-Ramond field. This is different from what occurs in MGD-decoupling, where the effect of $\Theta_{\mu\nu}$ can be neglected \cite{ovalle2018anisotropic}. Thus, the exterior spacetime manifold remains described by a vacuum spacetime.
  \item The analysis of the modified compactness factor $\frac{M}{R}$ reveals that the GR+MGD model admits more compact configurations than those that are permitted in the domain of pure GR. In the context of isotropic fluid distributions described by a monotonically decreasing energy density profile, Buchdahl \cite{buchdahl1959general} determined the upper bound for the mass-to-radius ratio. This limit is expressed as
     \begin{align}\label{w7}
\frac{M}{R}\leq \frac{4}{9}.
\end{align}
     Nevertheless, this upper limit can be exceeded in this case, with the new maximum being the black hole limit, $\frac{M}{R}= \frac{1}{2}$. Additionally, a lower bound for the compactness factor is acquired [108]. Naturally, all these factors depend on values of the deformation parameter and the compactness factor.
     \item Our findings indicate that MGD enhances both $M$ and $M/R$ as displayed in FIGs \ref{w2} and \ref{w3}. These modifications are significant from the perspective of cosmology and astrophysics. Through the application of MGD, the extension of isotropic distribution into an anisotropic domain can potentially have a higher mass density within the stellar system.
\end{itemize}}
\begin{figure*}[t]
\includegraphics{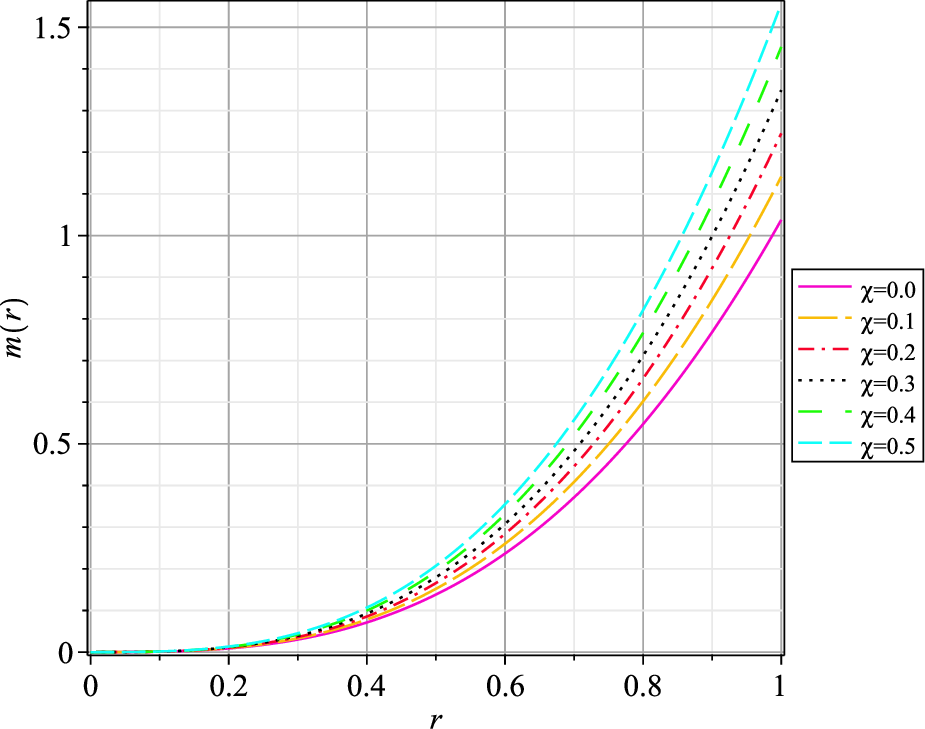}
\caption{\label{w2}The variation of mass function (m) versus $r$ for different $\chi$ values.}
\end{figure*}
\begin{figure*}[t]
\includegraphics{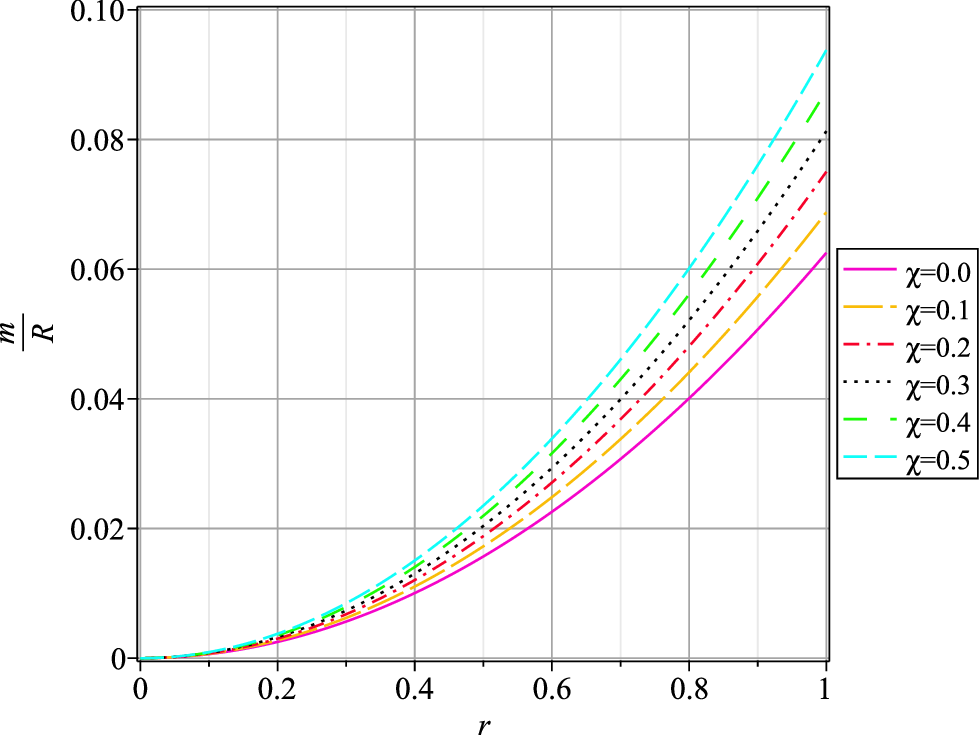}
\caption{\label{w3}The variation of mass-radius ratio ($M/R$) versus radial coordinate versus $r$ for different $\chi$ values.}
\end{figure*}

\section{\label{sec:level5}Energy conditions}

In addition, the considered matter distribution must satisfy some additional constraints to ensure its physical validity within the stellar interior. In this context, energy conditions act as constraints imposed on the SET $T_{\mu\nu}$, for modeling astrophysical dark stellar distributions. These limitations ensure the physically realistic behavior of matter content and prevent the emergence of unphysical energy distributions, often referred to as exotic matter \cite{ovalle2015brane}. In these stellar configurations, the energy density and pressure of the matter distributions often exhibit a linear dependence, subject to specific constraints. Therefore, a well-defined SET should satisfy the following inequalities
\begin{align}\label{m1}
&\textmd{NEC}:\quad\rho\geq0,
\\\label{q1}
&\textmd{WEC}:\quad\rho\geq0,~\rho+P_{r}\geq0,~\rho+P_{t}\geq0,
\\\label{m2}
&\textmd{SEC}:\quad\rho+P_{r}+2P_{t}\geq0,~\rho+P_{r}\geq0,~\rho+P_{t}\geq0,
\\\label{m3}
&\textmd{DEC}:\quad\rho\geq0,~\rho-|P_{r}|\geq0,~\rho-|P_{t}|\geq0,
\\\label{m4}
&\textmd{TEC}:\quad \rho-P_{r}-2P_{t}\geq0.
\end{align}
The physical interpretation of the energy conditions is defined as follows:
\begin{itemize}
  \item The null energy condition (NEC), weak energy condition (WEC), strong energy condition (SEC) dominant energy condition (DEC), and trace energy condition (TEC). These conditions collectively ensure that the observed energy density remains non-negative everywhere.
  \item The SEC is linked to the attractive nature of gravity. It imposes additional constraints on the SET that are consistent with gravitationally attractive forces.
  \item As observed in FIGs \ref{1f}-\ref{3f}, $\rho$, $P_{r}$, and $P_{t}$ all exhibit positive and finite values within the stellar region. This confirms the satisfaction of NEC and WEC in this domain.
  \item FIGs. \ref{7f} and \ref{8f} provide further validation by demonstrating that the DEC also holds throughout the interior of the star.
  \item It is evident from FIG. \ref{9f} that SEC is satisfied at all points, indicating that the core of a celestial body is characterized by a specified and positive SET.
  \item The TEC shows that $T_{\mu\nu}$ must be positive, as displayed in FIG. \ref{10f}.
\end{itemize}
The parameter $\chi$ directly affects all the inequalities outlined in the energy conditions, leading to increased positivity for higher $\chi$-values. Consequently, as the DM source strength intensifies, the energy conditions are more reliably satisfied within the minimally deformed Finch-Skea model.
Table~\ref{tab:tableS3} and Table~\ref{tab:tableS} describe the values of the energy conditions against the radial coordinate $r$ for various $\chi$ values.
\begin{table}[t]
\caption{\label{tab:tableS1}%
Values of the structural variables and the pressure anisotropy versus radial coordinate $r$ for different values of the decoupling parameter $\chi$.}

\begin{ruledtabular}
\begin{tabular}{cccccccc}
 &$r$&$\chi$&$\rho$&$P_{r}$&
 $P_{t}$&$\Pi$\\
\hline
1& 0.0 &0.0& 0.218 & 1.211 & 1.173 &0.000\\
2& 0.2 &0.1& 0.121 & 1.734 & 1.764
& 0.026\\
3& 0.4 &0.2& 0.193 & 1.315 & 1.328
& 0.077\\
4& 0.6 &0.3& 0.163 & 0.807 & 0.824
& 0.119\\
5& 0.8 &0.4& 0.137 & 0.359 & 0.402
& 0.136\\
6& 1.0 &0.5& 0.110 & 0.034 & 0.084
&  1.137\\
\end{tabular}
\end{ruledtabular}
\end{table}

\begin{table}
\caption{\label{tab:tableS3}%
Values of the NEC, WEC and DEC against the radial coordinate $r$ for various values of $\chi$.}
\begin{ruledtabular}
\begin{tabular}{cccccccc}
 &$r$&$\chi$&$\rho+P_{r}$&$\rho+P_{t}$&$\rho-P_{r}$&$\rho-P_{t}$\\
\hline
1& 0.0 &0.0& 1.408 & 1.43 & 1.591
& 1.408\\
2& 0.2 &0.1& 1.941 & 1.976 & 2.296
& 1.938\\
3& 0.4 &0.2& 1.530 & 0.532 & 1.754
& 1.531\\
 4& 0.6 &0.3& 0.963 & 0.005 & 1.045
& 0.964\\
5& 0.8 &0.4& 0.479 & 0.529 & 0.476
& 0.534\\
6& 1.0 &0.5& 0.118 & 0.155 & 0.044
&  0.192\\
\end{tabular}
\end{ruledtabular}
\end{table}

\begin{table}
\caption{\label{tab:tableS}%
Values of the SEC and TEC against the radial coordinate $r$ for different values of $\chi$.
}
\begin{ruledtabular}
\begin{tabular}{cccccccc}
 &$r$&$\chi$&$\rho+P_{r}+2P_{t}$&$\rho-P_{r}-2P_{t}$\\
\hline
1& 0.0& 0.0& 5.831 & 2.531 &
\\
2& 0.2& 0.1& 6.503 & 2.569 &
\\
3& 0.4& 0.2& 5.804 & 2.205 &
\\
4& 0.6& 0.3& 4.461 & 2.712 &
\\
5& 0.8& 0.4& 3.038 & 1.332 &
\\
6& 1.0& 0.5& 1.748 & 0.992 &
\\
\end{tabular}
\end{ruledtabular}
\end{table}
\section{\label{sec:level6}Conclusions}

Motivated by the DM dominance in our universe, we have constructed analytical solutions for an anisotropic fluid sphere based on a DM model. The metric of this compact configuration is described by a minimally deformed Adler-Finch-Skea spacetime. To achieve this, we employed a specific combination of the MGD decoupling and Class I approach, with the Einasto density profile serving as a key analytical tool.
To introduce anisotropic matter content into the isotropic fluid sphere, we employ gravitational decoupling via the MGD scheme. This scheme assumes deformation only in the radial metric potential, leaving the temporal metric potential unaltered. Within this framework, the temporal component, $\Theta^{t}_{t}$, of the decoupling fluid, $\Theta_{\mu\nu}$, is designed to mimic the Einasto DM density profile. The key ingredient of this study is the implementation of the Einasto DM density profile alongside the Class I condition. This combined approach serves as an alternative mimic constraint, effectively closing the relativistic system. To determine the deformation function $f(r)$, we supported our approach by applying it to a specific Class I metric solution: the Adler-Finch-Skea ansatz. In the context of gravitational decoupling, it is well understood that the $\Theta$-field source is fully determined once the temporal metric potential and the deformation function $f(r)$ are specified. This approach provides a novel method for identifying deformation functions $f(r)$ within the framework of gravitational decoupling. It builds upon previous works \cite{ovalle2018anisotropic,estrada2019gravitational,morales2018charged,gabbanelli2018gravitational}, where the mimic constraint scheme was employed to generate $f(r)$, or where a suitable $f(r)$ was imposed, as seen in \cite{maurya2019generalized,singh2019minimally}.
\begin{table*}
\caption{\label{tab:tableM}%
Summary of the work.}
\begin{ruledtabular}
\begin{tabular}{ccccc}
 Model &Density Profile&Geometric Deformation&Matter Variables
&Energy Conditions\\ \hline
 Adler-Finch-Skea &$\rho_{0}e^{-\left(\frac{r}{h}\right)^{1/n}}$&Minimal &$0<\rho,P_{r},P_{\bot}<\infty$&$\rho\geq0$ \\
 Adler-Finch-Skea &$\rho_{0}e^{-\left(\frac{r}{h}\right)^{1/n}}$
 &Minimal&$\Pi>0$&$\rho\geq0,~\rho+P_{r}\geq0,~\rho+P_{t}\geq0$\\
 Adler-Finch-Skea &$\rho_{0}e^{-\left(\frac{r}{h}\right)^{1/n}}$&Minimal
&$P_{r}=0 $~at~$r=R$&$\rho+P_{r}+2P_{t}\geq0,~\rho+P_{r}\geq0,~\rho+P_{t}\geq0$\\
 Adler-Finch-Skea &$\rho_{0}e^{-\left(\frac{r}{h}\right)^{1/n}}$&Minimal& decrease as $r$ increases&$\rho\geq0,~\rho-|P_{r}|\geq0,~\rho-|P_{t}|\geq0$\\
 Adler-Finch-Skea &$\rho_{0}e^{-\left(\frac{r}{h}\right)^{1/n}}$&Minimal& maximum at $r=0$&$\rho-P_{r}-2P_{t}\geq0$\\
\end{tabular}
\end{ruledtabular}
\end{table*}
This investigation explores the construction of stellar distributions with an anisotropic dark matter (DM) distribution using the non-commutativity-inspired Einasto density model. We show that coupling the Einasto profile with induced anisotropy through the MGD decoupling opens the possibility for various DM stellar structures. The Einasto parameterizations has been widely employed in modeling galactic structures with DM distribution. This generalizes the concept of non-commutative stellar configurations, such as black holes (BHs), and allows DM to enter as a matter constituent in self-gravitating compact systems. These types of stellar distributions could also be used to model the interior galactic structures surrounded by DM halos. Motivated by the DM dominance in the universe, this research also provides new insights into developing possible connections between DM and compact stars. Furthermore, the inclusion of the Einasto function as a DM profile can be used to model regular self-gravitational structures with fuzzy mass distribution as discussed in \cite{batic2021fuzzy,yousaf2024generating,khan2024structure}. We have calculated the values of the physical variables corresponding to the Einasto density profile by taking $n=0.7$, which corresponds to the DM case as described in the studies \cite{einasto1969galactic,einasto1969andromeda}

To ensure the physical validity of our results in representing realistic stars, we conducted a comprehensive analysis of the resulting compact object configurations, which incorporate anisotropic matter content. This analysis identifies all the essential requirements (or conditions) that a compact structure must meet to be considered admissible. Table~\ref{tab:tableS1} described the values of the matter variables against the radial variable and different values of $\chi$. Within this framework, we investigated the geometric structure and fundamental matter variables within the stellar distribution. Additionally, we examined the behavior of SET through the mechanism of MGD decoupling. Building upon the groundwork of MGD decoupling, we first analyzed the fundamental physical parameters for the spherically symmetric stellar object. These parameters include $\rho$, $P_{r}$, and $P_{t}$, as presented in FIGs. \ref{1f}-\ref{3f}. It is observed that $\rho$, $P_{r}$, and $P_{t}$ satisfy the essential requirements for a physically well-behaved compact system, i.e., (i) All three structural variables are positive and finite throughout the compact configuration; (ii) They display a peak at the center and decrease steadily as $r$ increases; iii) $P_{r}$ vanishes at the boundary. Furthermore, FIG. \ref{4f} (left panel) specifically confirms the positive nature of the anisotropic factor ($\Pi>0$). This property increases as $r$ increases, contributing to the stability of the self-gravitational compact system by maintaining hydrostatic equilibrium. While numerous key properties of the matter variables have already been addressed in conjunction with the corresponding figures (FIGs. \ref{1f}-\ref{4f}), we want to emphasize the critical role played by the decoupling constant $\chi$. Interestingly, $\rho$ exhibits an increase throughout the entire anisotropic dark star as the values of $\chi$ grow. Additionally, $P_{r}$ demonstrates a twofold behavior: it rises in the central region but converges near the stellar surface. The summary of the work is presented in Table~\ref{tab:tableM}.
Our findings demonstrate that the presented scheme serves as an efficient instrument for simulating compact objects within the framework of MGD decoupling in the presence of DM. We defer investigating the effects of DM on completely deformed astrophysical distributions within the Einasto density model to future studies.

\vspace{0.3cm}


\noindent {\bf Data Availability Statement:} This manuscript has no associated data or the data will not be deposited.

\vspace{0.5cm}

\noindent {\bf Declaration of Competing Interest:}
The authors declare that they have no known competing financial interests or personal relationships that could have appeared to influence the work reported in this paper.

\begin{acknowledgments}
The work of KB was supported by the JSPS KAKENHI Grant Numbers 21K03547, 23KF0008, and 24KF0100. The work by BA was supported by Researchers Supporting Project number: RSPD2024R650, King Saud University, Riyadh, Saudi Arabia.
\end{acknowledgments}

%


\begin{thebibliography}{104}%
\makeatletter
\providecommand \@ifxundefined [1]{%
 \@ifx{#1\undefined}
}%
\providecommand \@ifnum [1]{%
 \ifnum #1\expandafter \@firstoftwo
 \else \expandafter \@secondoftwo
 \fi
}%
\providecommand \@ifx [1]{%
 \ifx #1\expandafter \@firstoftwo
 \else \expandafter \@secondoftwo
 \fi
}%
\providecommand \natexlab [1]{#1}%
\providecommand \enquote  [1]{``#1''}%
\providecommand \bibnamefont  [1]{#1}%
\providecommand \bibfnamefont [1]{#1}%
\providecommand \citenamefont [1]{#1}%
\providecommand \href@noop [0]{\@secondoftwo}%
\providecommand \href [0]{\begingroup \@sanitize@url \@href}%
\providecommand \@href[1]{\@@startlink{#1}\@@href}%
\providecommand \@@href[1]{\endgroup#1\@@endlink}%
\providecommand \@sanitize@url [0]{\catcode `\\12\catcode `\$12\catcode `\&12\catcode `\#12\catcode `\^12\catcode `\_12\catcode `\%12\relax}%
\providecommand \@@startlink[1]{}%
\providecommand \@@endlink[0]{}%
\providecommand \url  [0]{\begingroup\@sanitize@url \@url }%
\providecommand \@url [1]{\endgroup\@href {#1}{\urlprefix }}%
\providecommand \urlprefix  [0]{URL }%
\providecommand \Eprint [0]{\href }%
\providecommand \doibase [0]{https://doi.org/}%
\providecommand \selectlanguage [0]{\@gobble}%
\providecommand \bibinfo  [0]{\@secondoftwo}%
\providecommand \bibfield  [0]{\@secondoftwo}%
\providecommand \translation [1]{[#1]}%
\providecommand \BibitemOpen [0]{}%
\providecommand \bibitemStop [0]{}%
\providecommand \bibitemNoStop [0]{.\EOS\space}%
\providecommand \EOS [0]{\spacefactor3000\relax}%
\providecommand \BibitemShut  [1]{\csname bibitem#1\endcsname}%
\let\auto@bib@innerbib\@empty
\bibitem [{\citenamefont {Jarosik}\ \emph {et~al.}(2011)\citenamefont {Jarosik} \emph {et~al.}}]{jarosik2011seven}%
  \BibitemOpen
  \bibfield  {author} {\bibinfo {author} {\bibfnamefont {N.}~\bibnamefont {Jarosik}} \emph {et~al.},\ }\bibfield  {title} {\bibinfo {title} {Seven-year wilkinson microwave anisotropy probe (\textsc{WMAP*}) observations: Sky maps, systematic errors, and basic results},\ }\href@noop {} {\bibfield  {journal} {\bibinfo  {journal} {Astrophys. J. Suppl. Ser.}\ }\textbf {\bibinfo {volume} {192}},\ \bibinfo {pages} {14} (\bibinfo {year} {2011})}\BibitemShut {NoStop}%
\bibitem [{\citenamefont {Ade}\ \emph {et~al.}(2014)\citenamefont {Ade} \emph {et~al.}}]{ade2014planck}%
  \BibitemOpen
  \bibfield  {author} {\bibinfo {author} {\bibfnamefont {P.~A.}\ \bibnamefont {Ade}} \emph {et~al.},\ }\bibfield  {title} {\bibinfo {title} {Planck 2013 results. \textsc{I}. overview of products and scientific results},\ }\href@noop {} {\bibfield  {journal} {\bibinfo  {journal} {Astron. Astrophys.}\ }\textbf {\bibinfo {volume} {571}},\ \bibinfo {pages} {A1} (\bibinfo {year} {2014})}\BibitemShut {NoStop}%
\bibitem [{\citenamefont {Faber}\ and\ \citenamefont {Gallagher}(1979)}]{faber1979masses}%
  \BibitemOpen
  \bibfield  {author} {\bibinfo {author} {\bibfnamefont {S.~M.}\ \bibnamefont {Faber}}\ and\ \bibinfo {author} {\bibfnamefont {J.}~\bibnamefont {Gallagher}},\ }\bibfield  {title} {\bibinfo {title} {Masses and mass-to-light ratios of galaxies},\ }\href@noop {} {\bibfield  {journal} {\bibinfo  {journal} {Annu. Rev. Astron. Astrophys.}\ }\textbf {\bibinfo {volume} {17}},\ \bibinfo {pages} {135} (\bibinfo {year} {1979})}\BibitemShut {NoStop}%
\bibitem [{\citenamefont {Bosma}(1981)}]{bosma198121}%
  \BibitemOpen
  \bibfield  {author} {\bibinfo {author} {\bibfnamefont {A.}~\bibnamefont {Bosma}},\ }\bibfield  {title} {\bibinfo {title} {21-cm line studies of spiral galaxies. \textsc{II}. the distribution and kinematics of neutral hydrogen in spiral galaxies of various morphological types.},\ }\href@noop {} {\bibfield  {journal} {\bibinfo  {journal} {Astron. J.}\ }\textbf {\bibinfo {volume} {86}},\ \bibinfo {pages} {1825} (\bibinfo {year} {1981})}\BibitemShut {NoStop}%
\bibitem [{\citenamefont {Barger}\ \emph {et~al.}(2002)\citenamefont {Barger}, \citenamefont {Halzen}, \citenamefont {Hooper},\ and\ \citenamefont {Kao}}]{barger2002indirect}%
  \BibitemOpen
  \bibfield  {author} {\bibinfo {author} {\bibfnamefont {V.}~\bibnamefont {Barger}}, \bibinfo {author} {\bibfnamefont {F.}~\bibnamefont {Halzen}}, \bibinfo {author} {\bibfnamefont {D.}~\bibnamefont {Hooper}},\ and\ \bibinfo {author} {\bibfnamefont {C.}~\bibnamefont {Kao}},\ }\bibfield  {title} {\bibinfo {title} {Indirect search for neutralino dark matter with high energy neutrinos},\ }\href@noop {} {\bibfield  {journal} {\bibinfo  {journal} {Phys. Rev. D}\ }\textbf {\bibinfo {volume} {65}},\ \bibinfo {pages} {075022} (\bibinfo {year} {2002})}\BibitemShut {NoStop}%
\bibitem [{\citenamefont {Spolyar}\ \emph {et~al.}(2008)\citenamefont {Spolyar}, \citenamefont {Freese},\ and\ \citenamefont {Gondolo}}]{spolyar2008dark}%
  \BibitemOpen
  \bibfield  {author} {\bibinfo {author} {\bibfnamefont {D.}~\bibnamefont {Spolyar}}, \bibinfo {author} {\bibfnamefont {K.}~\bibnamefont {Freese}},\ and\ \bibinfo {author} {\bibfnamefont {P.}~\bibnamefont {Gondolo}},\ }\bibfield  {title} {\bibinfo {title} {Dark matter and the first stars: a new phase of stellar evolution},\ }\href@noop {} {\bibfield  {journal} {\bibinfo  {journal} {Phys. Rev. Lett.}\ }\textbf {\bibinfo {volume} {100}},\ \bibinfo {pages} {051101} (\bibinfo {year} {2008})}\BibitemShut {NoStop}%
\bibitem [{\citenamefont {Hadjimichef}\ \emph {et~al.}(2017)\citenamefont {Hadjimichef}, \citenamefont {Machado}, \citenamefont {K{\"o}pp}, \citenamefont {Volkmer}, \citenamefont {Razeira},\ and\ \citenamefont {Vasconcellos}}]{hadjimichef2017dark}%
  \BibitemOpen
  \bibfield  {author} {\bibinfo {author} {\bibfnamefont {D.}~\bibnamefont {Hadjimichef}}, \bibinfo {author} {\bibfnamefont {M.}~\bibnamefont {Machado}}, \bibinfo {author} {\bibfnamefont {F.}~\bibnamefont {K{\"o}pp}}, \bibinfo {author} {\bibfnamefont {G.}~\bibnamefont {Volkmer}}, \bibinfo {author} {\bibfnamefont {M.}~\bibnamefont {Razeira}},\ and\ \bibinfo {author} {\bibfnamefont {C.}~\bibnamefont {Vasconcellos}},\ }\bibfield  {title} {\bibinfo {title} {A dark matter compact star in the framework of the pseudo-complex general relativity},\ }\href@noop {} {\bibfield  {journal} {\bibinfo  {journal} {Astron. Nachr.}\ }\textbf {\bibinfo {volume} {338}},\ \bibinfo {pages} {1079} (\bibinfo {year} {2017})}\BibitemShut {NoStop}%
\bibitem [{\citenamefont {Lopes}\ and\ \citenamefont {Panotopoulos}(2018)}]{lopes2018dark}%
  \BibitemOpen
  \bibfield  {author} {\bibinfo {author} {\bibfnamefont {I.}~\bibnamefont {Lopes}}\ and\ \bibinfo {author} {\bibfnamefont {G.}~\bibnamefont {Panotopoulos}},\ }\bibfield  {title} {\bibinfo {title} {Dark matter admixed strange quark stars in the starobinsky model},\ }\href@noop {} {\bibfield  {journal} {\bibinfo  {journal} {Phys. Rev. D}\ }\textbf {\bibinfo {volume} {97}},\ \bibinfo {pages} {024030} (\bibinfo {year} {2018})}\BibitemShut {NoStop}%
\bibitem [{\citenamefont {Rezaei}(2018{\natexlab{a}})}]{rezaei2018neutron}%
  \BibitemOpen
  \bibfield  {author} {\bibinfo {author} {\bibfnamefont {Z.}~\bibnamefont {Rezaei}},\ }\bibfield  {title} {\bibinfo {title} {Neutron stars with spin polarized self-interacting dark matter},\ }\href@noop {} {\bibfield  {journal} {\bibinfo  {journal} {Astroparticle Physics}\ }\textbf {\bibinfo {volume} {101}},\ \bibinfo {pages} {1} (\bibinfo {year} {2018}{\natexlab{a}})}\BibitemShut {NoStop}%
\bibitem [{\citenamefont {Rezaei}(2018{\natexlab{b}})}]{rezaei2018double}%
  \BibitemOpen
  \bibfield  {author} {\bibinfo {author} {\bibfnamefont {Z.}~\bibnamefont {Rezaei}},\ }\bibfield  {title} {\bibinfo {title} {Double dark matter admixed neutron star},\ }\href@noop {} {\bibfield  {journal} {\bibinfo  {journal} {Int. J. Mod. Phys. D}\ }\textbf {\bibinfo {volume} {27}},\ \bibinfo {pages} {1950002} (\bibinfo {year} {2018}{\natexlab{b}})}\BibitemShut {NoStop}%
\bibitem [{\citenamefont {Panotopoulos}\ and\ \citenamefont {Lopes}(2017)}]{panotopoulos2017gravitational}%
  \BibitemOpen
  \bibfield  {author} {\bibinfo {author} {\bibfnamefont {G.}~\bibnamefont {Panotopoulos}}\ and\ \bibinfo {author} {\bibfnamefont {I.}~\bibnamefont {Lopes}},\ }\bibfield  {title} {\bibinfo {title} {Gravitational effects of condensed dark matter on strange stars},\ }\href@noop {} {\bibfield  {journal} {\bibinfo  {journal} {Phys. Rev. D}\ }\textbf {\bibinfo {volume} {96}},\ \bibinfo {pages} {023002} (\bibinfo {year} {2017})}\BibitemShut {NoStop}%
\bibitem [{\citenamefont {Li}\ \emph {et~al.}(2012)\citenamefont {Li}, \citenamefont {Wang},\ and\ \citenamefont {Cheng}}]{li2012gravitational}%
  \BibitemOpen
  \bibfield  {author} {\bibinfo {author} {\bibfnamefont {X.}~\bibnamefont {Li}}, \bibinfo {author} {\bibfnamefont {F.}~\bibnamefont {Wang}},\ and\ \bibinfo {author} {\bibfnamefont {K.}~\bibnamefont {Cheng}},\ }\bibfield  {title} {\bibinfo {title} {Gravitational effects of condensate dark matter on compact stellar objects},\ }\href@noop {} {\bibfield  {journal} {\bibinfo  {journal} {J. Cosmol. Astropart. Phys.}\ }\textbf {\bibinfo {volume} {2012}},\ \bibinfo {pages} {031}}\BibitemShut {NoStop}%
\bibitem [{\citenamefont {Ciarcelluti}\ and\ \citenamefont {Sandin}(2011)}]{ciarcelluti2011have}%
  \BibitemOpen
  \bibfield  {author} {\bibinfo {author} {\bibfnamefont {P.}~\bibnamefont {Ciarcelluti}}\ and\ \bibinfo {author} {\bibfnamefont {F.}~\bibnamefont {Sandin}},\ }\bibfield  {title} {\bibinfo {title} {Have neutron stars a dark matter core?},\ }\href@noop {} {\bibfield  {journal} {\bibinfo  {journal} {Phys. Lett. B}\ }\textbf {\bibinfo {volume} {695}},\ \bibinfo {pages} {19} (\bibinfo {year} {2011})}\BibitemShut {NoStop}%
\bibitem [{\citenamefont {Delgaty}\ and\ \citenamefont {Lake}(1998)}]{delgaty1998physical}%
  \BibitemOpen
  \bibfield  {author} {\bibinfo {author} {\bibfnamefont {M.}~\bibnamefont {Delgaty}}\ and\ \bibinfo {author} {\bibfnamefont {K.}~\bibnamefont {Lake}},\ }\bibfield  {title} {\bibinfo {title} {Physical acceptability of isolated, static, spherically symmetric, perfect fluid solutions of einstein's equations},\ }\href@noop {} {\bibfield  {journal} {\bibinfo  {journal} {Comput. Phys. Commun.}\ }\textbf {\bibinfo {volume} {115}},\ \bibinfo {pages} {395} (\bibinfo {year} {1998})}\BibitemShut {NoStop}%
\bibitem [{\citenamefont {Herrera}\ and\ \citenamefont {Santos}(1997)}]{herrera1997local}%
  \BibitemOpen
  \bibfield  {author} {\bibinfo {author} {\bibfnamefont {L.}~\bibnamefont {Herrera}}\ and\ \bibinfo {author} {\bibfnamefont {N.~O.}\ \bibnamefont {Santos}},\ }\bibfield  {title} {\bibinfo {title} {Local anisotropy in self-gravitating systems},\ }\href@noop {} {\bibfield  {journal} {\bibinfo  {journal} {Phys. Rep.}\ }\textbf {\bibinfo {volume} {286}},\ \bibinfo {pages} {53} (\bibinfo {year} {1997})}\BibitemShut {NoStop}%
\bibitem [{\citenamefont {Ruderman}(1972)}]{ruderman1972pulsars}%
  \BibitemOpen
  \bibfield  {author} {\bibinfo {author} {\bibfnamefont {M.}~\bibnamefont {Ruderman}},\ }\bibfield  {title} {\bibinfo {title} {Pulsars: structure and dynamics},\ }\href@noop {} {\bibfield  {journal} {\bibinfo  {journal} {Ann. Rev. Astron. Astrophys}\ }\textbf {\bibinfo {volume} {10}},\ \bibinfo {pages} {427} (\bibinfo {year} {1972})}\BibitemShut {NoStop}%
\bibitem [{\citenamefont {Poplawski}(2013{\natexlab{a}})}]{poplawski2013intrinsic}%
  \BibitemOpen
  \bibfield  {author} {\bibinfo {author} {\bibfnamefont {N.}~\bibnamefont {Poplawski}},\ }\bibfield  {title} {\bibinfo {title} {Intrinsic spin requires gravity with torsion and curvature},\ }\href@noop {} {\bibfield  {journal} {\bibinfo  {journal} {arXiv preprint arXiv:1304.0047}\ } (\bibinfo {year} {2013}{\natexlab{a}})}\BibitemShut {NoStop}%
\bibitem [{\citenamefont {Pop{\l}awski}(2021)}]{poplawski2021gravitational}%
  \BibitemOpen
  \bibfield  {author} {\bibinfo {author} {\bibfnamefont {N.}~\bibnamefont {Pop{\l}awski}},\ }\bibfield  {title} {\bibinfo {title} {Gravitational collapse of a fluid with torsion into a universe in a black hole},\ }\href@noop {} {\bibfield  {journal} {\bibinfo  {journal} {J. Exp. Theor. Phys.}\ }\textbf {\bibinfo {volume} {132}},\ \bibinfo {pages} {374} (\bibinfo {year} {2021})}\BibitemShut {NoStop}%
\bibitem [{\citenamefont {Pop{\l}awski}(2012)}]{poplawski2012four}%
  \BibitemOpen
  \bibfield  {author} {\bibinfo {author} {\bibfnamefont {N.~J.}\ \bibnamefont {Pop{\l}awski}},\ }\bibfield  {title} {\bibinfo {title} {Four-fermion interaction from torsion as dark energy},\ }\href@noop {} {\bibfield  {journal} {\bibinfo  {journal} {Gen. Relativ. Gravitation}\ }\textbf {\bibinfo {volume} {44}},\ \bibinfo {pages} {491} (\bibinfo {year} {2012})}\BibitemShut {NoStop}%
\bibitem [{\citenamefont {Poplawski}(2013{\natexlab{b}})}]{poplawski2013cosmological}%
  \BibitemOpen
  \bibfield  {author} {\bibinfo {author} {\bibfnamefont {N.~J.}\ \bibnamefont {Poplawski}},\ }\bibfield  {title} {\bibinfo {title} {Cosmological consequences of gravity with spin and torsion},\ }\href@noop {} {\bibfield  {journal} {\bibinfo  {journal} {Astron. Rev.}\ }\textbf {\bibinfo {volume} {8}},\ \bibinfo {pages} {108} (\bibinfo {year} {2013}{\natexlab{b}})}\BibitemShut {NoStop}%
\bibitem [{\citenamefont {Khlopov}\ \emph {et~al.}(1985)\citenamefont {Khlopov}, \citenamefont {Malomed},\ and\ \citenamefont {Zeldovich}}]{khlopov1985gravitational}%
  \BibitemOpen
  \bibfield  {author} {\bibinfo {author} {\bibfnamefont {M.~Y.}\ \bibnamefont {Khlopov}}, \bibinfo {author} {\bibfnamefont {B.}~\bibnamefont {Malomed}},\ and\ \bibinfo {author} {\bibfnamefont {Y.~B.}\ \bibnamefont {Zeldovich}},\ }\bibfield  {title} {\bibinfo {title} {Gravitational instability of scalar fields and formation of primordial black holes},\ }\href@noop {} {\bibfield  {journal} {\bibinfo  {journal} {Mon. Not. R. Astron. Soc.}\ }\textbf {\bibinfo {volume} {215}},\ \bibinfo {pages} {575} (\bibinfo {year} {1985})}\BibitemShut {NoStop}%
\bibitem [{\citenamefont {Di~Grezia}\ \emph {et~al.}(2017)\citenamefont {Di~Grezia}, \citenamefont {Battista}, \citenamefont {Manfredonia},\ and\ \citenamefont {Miele}}]{di2017spin}%
  \BibitemOpen
  \bibfield  {author} {\bibinfo {author} {\bibfnamefont {E.}~\bibnamefont {Di~Grezia}}, \bibinfo {author} {\bibfnamefont {E.}~\bibnamefont {Battista}}, \bibinfo {author} {\bibfnamefont {M.}~\bibnamefont {Manfredonia}},\ and\ \bibinfo {author} {\bibfnamefont {G.}~\bibnamefont {Miele}},\ }\bibfield  {title} {\bibinfo {title} {Spin, torsion and violation of null energy condition in traversable wormholes},\ }\href@noop {} {\bibfield  {journal} {\bibinfo  {journal} {Eur. Phys. J. Plus}\ }\textbf {\bibinfo {volume} {132}},\ \bibinfo {pages} {537} (\bibinfo {year} {2017})}\BibitemShut {NoStop}%
\bibitem [{\citenamefont {De~Falco}\ \emph {et~al.}(2020)\citenamefont {De~Falco}, \citenamefont {Battista}, \citenamefont {Capozziello},\ and\ \citenamefont {De~Laurentis}}]{de2020general}%
  \BibitemOpen
  \bibfield  {author} {\bibinfo {author} {\bibfnamefont {V.}~\bibnamefont {De~Falco}}, \bibinfo {author} {\bibfnamefont {E.}~\bibnamefont {Battista}}, \bibinfo {author} {\bibfnamefont {S.}~\bibnamefont {Capozziello}},\ and\ \bibinfo {author} {\bibfnamefont {M.}~\bibnamefont {De~Laurentis}},\ }\bibfield  {title} {\bibinfo {title} {General relativistic poynting-robertson effect to diagnose wormholes existence: Static and spherically symmetric case},\ }\href@noop {} {\bibfield  {journal} {\bibinfo  {journal} {Phys. Rev. D}\ }\textbf {\bibinfo {volume} {101}},\ \bibinfo {pages} {104037} (\bibinfo {year} {2020})}\BibitemShut {NoStop}%
\bibitem [{\citenamefont {De~Falco}\ \emph {et~al.}(2021)\citenamefont {De~Falco}, \citenamefont {Battista}, \citenamefont {Capozziello},\ and\ \citenamefont {De~Laurentis}}]{de2021testing}%
  \BibitemOpen
  \bibfield  {author} {\bibinfo {author} {\bibfnamefont {V.}~\bibnamefont {De~Falco}}, \bibinfo {author} {\bibfnamefont {E.}~\bibnamefont {Battista}}, \bibinfo {author} {\bibfnamefont {S.}~\bibnamefont {Capozziello}},\ and\ \bibinfo {author} {\bibfnamefont {M.}~\bibnamefont {De~Laurentis}},\ }\bibfield  {title} {\bibinfo {title} {Testing wormhole solutions in extended gravity through the poynting-robertson effect},\ }\href@noop {} {\bibfield  {journal} {\bibinfo  {journal} {Phys. Rev. D}\ }\textbf {\bibinfo {volume} {103}},\ \bibinfo {pages} {044007} (\bibinfo {year} {2021})}\BibitemShut {NoStop}%
\bibitem [{\citenamefont {Astashenok}\ and\ \citenamefont {Odintsov}(2020)}]{astashenok2020supermassive}%
  \BibitemOpen
  \bibfield  {author} {\bibinfo {author} {\bibfnamefont {A.~V.}\ \bibnamefont {Astashenok}}\ and\ \bibinfo {author} {\bibfnamefont {S.~D.}\ \bibnamefont {Odintsov}},\ }\bibfield  {title} {\bibinfo {title} {Supermassive neutron stars in axion ${F (R)}$ gravity},\ }\href@noop {} {\bibfield  {journal} {\bibinfo  {journal} {Mon. Not. R. Astron. Soc.}\ }\textbf {\bibinfo {volume} {493}},\ \bibinfo {pages} {78} (\bibinfo {year} {2020})}\BibitemShut {NoStop}%
\bibitem [{\citenamefont {Nojiri}\ \emph {et~al.}(2020)\citenamefont {Nojiri}, \citenamefont {Odintsov},\ and\ \citenamefont {Oikonomou}}]{nojiri2020f}%
  \BibitemOpen
  \bibfield  {author} {\bibinfo {author} {\bibfnamefont {S.}~\bibnamefont {Nojiri}}, \bibinfo {author} {\bibfnamefont {S.~D.}\ \bibnamefont {Odintsov}},\ and\ \bibinfo {author} {\bibfnamefont {V.~K.}\ \bibnamefont {Oikonomou}},\ }\bibfield  {title} {\bibinfo {title} {${F (R)}$ gravity with an axion-like particle: Dynamics, gravity waves, late and early-time phenomenology},\ }\href@noop {} {\bibfield  {journal} {\bibinfo  {journal} {Ann. Phys.}\ }\textbf {\bibinfo {volume} {418}},\ \bibinfo {pages} {168186} (\bibinfo {year} {2020})}\BibitemShut {NoStop}%
\bibitem [{\citenamefont {Bowers}\ and\ \citenamefont {Liang}(1974)}]{bowers1974anisotropic}%
  \BibitemOpen
  \bibfield  {author} {\bibinfo {author} {\bibfnamefont {R.~L.}\ \bibnamefont {Bowers}}\ and\ \bibinfo {author} {\bibfnamefont {E.}~\bibnamefont {Liang}},\ }\bibfield  {title} {\bibinfo {title} {Anisotropic spheres in general relativity},\ }\href@noop {} {\bibfield  {journal} {\bibinfo  {journal} {Astrophys. J.}\ }\textbf {\bibinfo {volume} {188}},\ \bibinfo {pages} {657} (\bibinfo {year} {1974})}\BibitemShut {NoStop}%
\bibitem [{\citenamefont {Heintzmann}\ and\ \citenamefont {Hillebrandt}(1975)}]{heintzmann1975neutron}%
  \BibitemOpen
  \bibfield  {author} {\bibinfo {author} {\bibfnamefont {H.}~\bibnamefont {Heintzmann}}\ and\ \bibinfo {author} {\bibfnamefont {W.}~\bibnamefont {Hillebrandt}},\ }\bibfield  {title} {\bibinfo {title} {Neutron stars with an anisotropic equation of state-mass, redshift and stability},\ }\href@noop {} {\bibfield  {journal} {\bibinfo  {journal} {Astron. Astrophys.}\ }\textbf {\bibinfo {volume} {38}},\ \bibinfo {pages} {51} (\bibinfo {year} {1975})}\BibitemShut {NoStop}%
\bibitem [{\citenamefont {Cosenza}\ \emph {et~al.}(1982)\citenamefont {Cosenza}, \citenamefont {Herrera}, \citenamefont {Esculpi},\ and\ \citenamefont {Witten}}]{cosenza1982evolution}%
  \BibitemOpen
  \bibfield  {author} {\bibinfo {author} {\bibfnamefont {M.}~\bibnamefont {Cosenza}}, \bibinfo {author} {\bibfnamefont {L.}~\bibnamefont {Herrera}}, \bibinfo {author} {\bibfnamefont {M.}~\bibnamefont {Esculpi}},\ and\ \bibinfo {author} {\bibfnamefont {L.}~\bibnamefont {Witten}},\ }\bibfield  {title} {\bibinfo {title} {Evolution of radiating anisotropic spheres in general relativity},\ }\href@noop {} {\bibfield  {journal} {\bibinfo  {journal} {Phys. Rev. D}\ }\textbf {\bibinfo {volume} {25}},\ \bibinfo {pages} {2527} (\bibinfo {year} {1982})}\BibitemShut {NoStop}%
\bibitem [{\citenamefont {Herrera}\ and\ \citenamefont {Ponce~de Leon}(1985)}]{herrera1985isotropic}%
  \BibitemOpen
  \bibfield  {author} {\bibinfo {author} {\bibfnamefont {L.}~\bibnamefont {Herrera}}\ and\ \bibinfo {author} {\bibfnamefont {J.}~\bibnamefont {Ponce~de Leon}},\ }\bibfield  {title} {\bibinfo {title} {Isotropic and anisotropic charged spheres admitting a one-parameter group of conformal motions},\ }\href@noop {} {\bibfield  {journal} {\bibinfo  {journal} {J. Math. Phys.}\ }\textbf {\bibinfo {volume} {26}},\ \bibinfo {pages} {2302} (\bibinfo {year} {1985})}\BibitemShut {NoStop}%
\bibitem [{\citenamefont {Chan}\ \emph {et~al.}(1993)\citenamefont {Chan}, \citenamefont {Herrera},\ and\ \citenamefont {Santos}}]{chan1993dynamical}%
  \BibitemOpen
  \bibfield  {author} {\bibinfo {author} {\bibfnamefont {R.}~\bibnamefont {Chan}}, \bibinfo {author} {\bibfnamefont {L.}~\bibnamefont {Herrera}},\ and\ \bibinfo {author} {\bibfnamefont {N.}~\bibnamefont {Santos}},\ }\bibfield  {title} {\bibinfo {title} {Dynamical instability for radiating anisotropic collapse},\ }\href@noop {} {\bibfield  {journal} {\bibinfo  {journal} {Mon. Not. R. Astron. Soc.}\ }\textbf {\bibinfo {volume} {265}},\ \bibinfo {pages} {533} (\bibinfo {year} {1993})}\BibitemShut {NoStop}%
\bibitem [{\citenamefont {Di~Prisco}\ \emph {et~al.}(1997)\citenamefont {Di~Prisco}, \citenamefont {Herrera},\ and\ \citenamefont {Varela}}]{di1997cracking}%
  \BibitemOpen
  \bibfield  {author} {\bibinfo {author} {\bibfnamefont {A.}~\bibnamefont {Di~Prisco}}, \bibinfo {author} {\bibfnamefont {L.}~\bibnamefont {Herrera}},\ and\ \bibinfo {author} {\bibfnamefont {V.}~\bibnamefont {Varela}},\ }\bibfield  {title} {\bibinfo {title} {Cracking of homogeneous self-gravitating compact objects induced by fluctuations of local anisotropy},\ }\href@noop {} {\bibfield  {journal} {\bibinfo  {journal} {Gen. Relativ. Gravit.}\ }\textbf {\bibinfo {volume} {29}},\ \bibinfo {pages} {1239} (\bibinfo {year} {1997})}\BibitemShut {NoStop}%
\bibitem [{\citenamefont {Dev}\ and\ \citenamefont {Gleiser}(2002)}]{dev2002anisotropic}%
  \BibitemOpen
  \bibfield  {author} {\bibinfo {author} {\bibfnamefont {K.}~\bibnamefont {Dev}}\ and\ \bibinfo {author} {\bibfnamefont {M.}~\bibnamefont {Gleiser}},\ }\bibfield  {title} {\bibinfo {title} {Anisotropic stars: exact solutions},\ }\href@noop {} {\bibfield  {journal} {\bibinfo  {journal} {Gen. Relativ. Gravit.}\ }\textbf {\bibinfo {volume} {34}},\ \bibinfo {pages} {1793} (\bibinfo {year} {2002})}\BibitemShut {NoStop}%
\bibitem [{\citenamefont {Yousaf}\ \emph {et~al.}(2022)\citenamefont {Yousaf}, \citenamefont {Bhatti}, \citenamefont {Khan},\ and\ \citenamefont {Sahoo}}]{yousaf2022f}%
  \BibitemOpen
  \bibfield  {author} {\bibinfo {author} {\bibfnamefont {Z.}~\bibnamefont {Yousaf}}, \bibinfo {author} {\bibfnamefont {M.~Z.}\ \bibnamefont {Bhatti}}, \bibinfo {author} {\bibfnamefont {S.}~\bibnamefont {Khan}},\ and\ \bibinfo {author} {\bibfnamefont {P.~K.}\ \bibnamefont {Sahoo}},\ }\bibfield  {title} {\bibinfo {title} {${f(G, T_{\alpha\beta}T^{\alpha\beta})}$ theory and complex cosmological structures},\ }\href@noop {} {\bibfield  {journal} {\bibinfo  {journal} {Phys. Dark Universe}\ }\textbf {\bibinfo {volume} {36}},\ \bibinfo {pages} {101015} (\bibinfo {year} {2022})}\BibitemShut {NoStop}%
\bibitem [{\citenamefont {Malik}\ \emph {et~al.}(2024{\natexlab{a}})\citenamefont {Malik}, \citenamefont {Arif},\ and\ \citenamefont {Shamir}}]{malik2024charged}%
  \BibitemOpen
  \bibfield  {author} {\bibinfo {author} {\bibfnamefont {A.}~\bibnamefont {Malik}}, \bibinfo {author} {\bibfnamefont {A.}~\bibnamefont {Arif}},\ and\ \bibinfo {author} {\bibfnamefont {M.~F.}\ \bibnamefont {Shamir}},\ }\bibfield  {title} {\bibinfo {title} {Charged anisotropic compact stars in \textsc{R}icci-inverse gravity},\ }\href@noop {} {\bibfield  {journal} {\bibinfo  {journal} {Eur. Phys. J. Plus}\ }\textbf {\bibinfo {volume} {139}},\ \bibinfo {pages} {67} (\bibinfo {year} {2024}{\natexlab{a}})}\BibitemShut {NoStop}%
\bibitem [{\citenamefont {Malik}\ \emph {et~al.}(2024{\natexlab{b}})\citenamefont {Malik}, \citenamefont {Naz}, \citenamefont {Yousaf}, \citenamefont {Almas},\ and\ \citenamefont {Saleem}}]{malik2024singularity}%
  \BibitemOpen
  \bibfield  {author} {\bibinfo {author} {\bibfnamefont {A.}~\bibnamefont {Malik}}, \bibinfo {author} {\bibfnamefont {T.}~\bibnamefont {Naz}}, \bibinfo {author} {\bibfnamefont {Z.}~\bibnamefont {Yousaf}}, \bibinfo {author} {\bibfnamefont {A.}~\bibnamefont {Almas}},\ and\ \bibinfo {author} {\bibfnamefont {K.}~\bibnamefont {Saleem}},\ }\bibfield  {title} {\bibinfo {title} {Singularity-free anisotropic compact star in ${f(R,\phi)}$ gravity via \textsc{K}armarkar condition},\ }\href@noop {} {\bibfield  {journal} {\bibinfo  {journal} {Int. J. Geom. Methods Mod. Phys.}\ }\textbf {\bibinfo {volume} {21}},\ \bibinfo {pages} {2450018} (\bibinfo {year} {2024}{\natexlab{b}})}\BibitemShut {NoStop}%
\bibitem [{\citenamefont {Sokolov}(1980)}]{sokolov1980phase}%
  \BibitemOpen
  \bibfield  {author} {\bibinfo {author} {\bibfnamefont {A.}~\bibnamefont {Sokolov}},\ }\bibfield  {title} {\bibinfo {title} {Phase transformations in a superfluid neutron liquid},\ }\href@noop {} {\bibfield  {journal} {\bibinfo  {journal} {Zhurnal Ehksperimental'noj i Teoreticheskoj Fiziki}\ }\textbf {\bibinfo {volume} {49}},\ \bibinfo {pages} {1137} (\bibinfo {year} {1980})}\BibitemShut {NoStop}%
\bibitem [{\citenamefont {Sawyer}(1972)}]{sawyer1972condensed}%
  \BibitemOpen
  \bibfield  {author} {\bibinfo {author} {\bibfnamefont {R.~F.}\ \bibnamefont {Sawyer}},\ }\bibfield  {title} {\bibinfo {title} {Condensed $\pi$- phase in neutron-star matter},\ }\href@noop {} {\bibfield  {journal} {\bibinfo  {journal} {Phys. Rev. Lett.}\ }\textbf {\bibinfo {volume} {29}},\ \bibinfo {pages} {382} (\bibinfo {year} {1972})}\BibitemShut {NoStop}%
\bibitem [{\citenamefont {Albalahi}\ \emph {et~al.}(2024{\natexlab{a}})\citenamefont {Albalahi}, \citenamefont {Bhatti}, \citenamefont {Ali},\ and\ \citenamefont {Khan}}]{albalahi2024electromagnetic}%
  \BibitemOpen
  \bibfield  {author} {\bibinfo {author} {\bibfnamefont {A.~M.}\ \bibnamefont {Albalahi}}, \bibinfo {author} {\bibfnamefont {M.}~\bibnamefont {Bhatti}}, \bibinfo {author} {\bibfnamefont {A.}~\bibnamefont {Ali}},\ and\ \bibinfo {author} {\bibfnamefont {S.}~\bibnamefont {Khan}},\ }\bibfield  {title} {\bibinfo {title} {Electromagnetic field on the complexity of minimally deformed compact stars},\ }\href@noop {} {\bibfield  {journal} {\bibinfo  {journal} {Eur. Phys. J. C}\ }\textbf {\bibinfo {volume} {84}},\ \bibinfo {pages} {293} (\bibinfo {year} {2024}{\natexlab{a}})}\BibitemShut {NoStop}%
\bibitem [{\citenamefont {Albalahi}\ \emph {et~al.}(2024{\natexlab{b}})\citenamefont {Albalahi}, \citenamefont {Yousaf}, \citenamefont {Ali},\ and\ \citenamefont {Khan}}]{albalahi2024isotropization}%
  \BibitemOpen
  \bibfield  {author} {\bibinfo {author} {\bibfnamefont {A.~M.}\ \bibnamefont {Albalahi}}, \bibinfo {author} {\bibfnamefont {Z.}~\bibnamefont {Yousaf}}, \bibinfo {author} {\bibfnamefont {A.}~\bibnamefont {Ali}},\ and\ \bibinfo {author} {\bibfnamefont {S.}~\bibnamefont {Khan}},\ }\bibfield  {title} {\bibinfo {title} {Isotropization and complexity shift of gravitationally decoupled charged anisotropic sources},\ }\href@noop {} {\bibfield  {journal} {\bibinfo  {journal} {Eur. Phys. J. C}\ }\textbf {\bibinfo {volume} {84}},\ \bibinfo {pages} {9} (\bibinfo {year} {2024}{\natexlab{b}})}\BibitemShut {NoStop}%
\bibitem [{\citenamefont {Yousaf}\ \emph {et~al.}(2024{\natexlab{a}})\citenamefont {Yousaf}, \citenamefont {Khan}, \citenamefont {Turki},\ and\ \citenamefont {Suzuki}}]{yousaf2024modeling}%
  \BibitemOpen
  \bibfield  {author} {\bibinfo {author} {\bibfnamefont {Z.}~\bibnamefont {Yousaf}}, \bibinfo {author} {\bibfnamefont {S.}~\bibnamefont {Khan}}, \bibinfo {author} {\bibfnamefont {N.~B.}\ \bibnamefont {Turki}},\ and\ \bibinfo {author} {\bibfnamefont {T.}~\bibnamefont {Suzuki}},\ }\bibfield  {title} {\bibinfo {title} {Modeling of self-gravitating compact configurations using radial metric deformation approach},\ }\href@noop {} {\bibfield  {journal} {\bibinfo  {journal} {Chinese J. Phys.}\ }\textbf {\bibinfo {volume} {89}},\ \bibinfo {pages} {1595} (\bibinfo {year} {2024}{\natexlab{a}})}\BibitemShut {NoStop}%
\bibitem [{\citenamefont {Maurya}\ \emph {et~al.}(2016)\citenamefont {Maurya}, \citenamefont {Jasim}, \citenamefont {Gupta},\ and\ \citenamefont {Smitha}}]{maurya2016relativistic}%
  \BibitemOpen
  \bibfield  {author} {\bibinfo {author} {\bibfnamefont {S.}~\bibnamefont {Maurya}}, \bibinfo {author} {\bibfnamefont {M.}~\bibnamefont {Jasim}}, \bibinfo {author} {\bibfnamefont {Y.}~\bibnamefont {Gupta}},\ and\ \bibinfo {author} {\bibfnamefont {T.}~\bibnamefont {Smitha}},\ }\bibfield  {title} {\bibinfo {title} {Relativistic polytropic models for neutral stars with vanishing pressure anisotropy},\ }\href@noop {} {\bibfield  {journal} {\bibinfo  {journal} {Astrophys. Space Sci.}\ }\textbf {\bibinfo {volume} {361}},\ \bibinfo {pages} {163} (\bibinfo {year} {2016})}\BibitemShut {NoStop}%
\bibitem [{\citenamefont {Germani}\ and\ \citenamefont {Maartens}(2001)}]{germani2001stars}%
  \BibitemOpen
  \bibfield  {author} {\bibinfo {author} {\bibfnamefont {C.}~\bibnamefont {Germani}}\ and\ \bibinfo {author} {\bibfnamefont {R.}~\bibnamefont {Maartens}},\ }\bibfield  {title} {\bibinfo {title} {Stars in the braneworld},\ }\href@noop {} {\bibfield  {journal} {\bibinfo  {journal} {Phys. Rev. D}\ }\textbf {\bibinfo {volume} {64}},\ \bibinfo {pages} {124010} (\bibinfo {year} {2001})}\BibitemShut {NoStop}%
\bibitem [{\citenamefont {Ovalle}(2008)}]{ovalle2008searching}%
  \BibitemOpen
  \bibfield  {author} {\bibinfo {author} {\bibfnamefont {J.}~\bibnamefont {Ovalle}},\ }\bibfield  {title} {\bibinfo {title} {Searching exact solutions for compact stars in braneworld: a conjecture},\ }\href@noop {} {\bibfield  {journal} {\bibinfo  {journal} {Mod. Phys. Lett. A}\ }\textbf {\bibinfo {volume} {23}},\ \bibinfo {pages} {3247} (\bibinfo {year} {2008})}\BibitemShut {NoStop}%
\bibitem [{\citenamefont {Ovalle}\ \emph {et~al.}(2015)\citenamefont {Ovalle}, \citenamefont {Gergely},\ and\ \citenamefont {Casadio}}]{ovalle2015brane}%
  \BibitemOpen
  \bibfield  {author} {\bibinfo {author} {\bibfnamefont {J.}~\bibnamefont {Ovalle}}, \bibinfo {author} {\bibfnamefont {L.~A.}\ \bibnamefont {Gergely}},\ and\ \bibinfo {author} {\bibfnamefont {R.}~\bibnamefont {Casadio}},\ }\bibfield  {title} {\bibinfo {title} {Brane-world stars with a solid crust and vacuum exterior},\ }\href@noop {} {\bibfield  {journal} {\bibinfo  {journal} {Class. Quantum Grav.}\ }\textbf {\bibinfo {volume} {32}},\ \bibinfo {pages} {045015} (\bibinfo {year} {2015})}\BibitemShut {NoStop}%
\bibitem [{\citenamefont {Ovalle}(2017)}]{ovalle2017decoupling}%
  \BibitemOpen
  \bibfield  {author} {\bibinfo {author} {\bibfnamefont {J.}~\bibnamefont {Ovalle}},\ }\bibfield  {title} {\bibinfo {title} {Decoupling gravitational sources in general relativity: from perfect to anisotropic fluids},\ }\href@noop {} {\bibfield  {journal} {\bibinfo  {journal} {Phys. Rev. D}\ }\textbf {\bibinfo {volume} {95}},\ \bibinfo {pages} {104019} (\bibinfo {year} {2017})}\BibitemShut {NoStop}%
\bibitem [{\citenamefont {Ovalle}\ \emph {et~al.}(2018{\natexlab{a}})\citenamefont {Ovalle}, \citenamefont {Casadio}, \citenamefont {Da~Rocha},\ and\ \citenamefont {Sotomayor}}]{ovalle2018anisotropic}%
  \BibitemOpen
  \bibfield  {author} {\bibinfo {author} {\bibfnamefont {J.}~\bibnamefont {Ovalle}}, \bibinfo {author} {\bibfnamefont {R.}~\bibnamefont {Casadio}}, \bibinfo {author} {\bibfnamefont {R.}~\bibnamefont {Da~Rocha}},\ and\ \bibinfo {author} {\bibfnamefont {A.}~\bibnamefont {Sotomayor}},\ }\bibfield  {title} {\bibinfo {title} {Anisotropic solutions by gravitational decoupling},\ }\href@noop {} {\bibfield  {journal} {\bibinfo  {journal} {Eur. Phys. J. C}\ }\textbf {\bibinfo {volume} {78}},\ \bibinfo {pages} {122} (\bibinfo {year} {2018}{\natexlab{a}})}\BibitemShut {NoStop}%
\bibitem [{\citenamefont {Casadio}\ \emph {et~al.}(2015)\citenamefont {Casadio}, \citenamefont {Ovalle},\ and\ \citenamefont {Da~Rocha}}]{casadio2015minimal}%
  \BibitemOpen
  \bibfield  {author} {\bibinfo {author} {\bibfnamefont {R.}~\bibnamefont {Casadio}}, \bibinfo {author} {\bibfnamefont {J.}~\bibnamefont {Ovalle}},\ and\ \bibinfo {author} {\bibfnamefont {R.}~\bibnamefont {Da~Rocha}},\ }\bibfield  {title} {\bibinfo {title} {The minimal geometric deformation approach extended},\ }\href@noop {} {\bibfield  {journal} {\bibinfo  {journal} {Class. Quantum Grav.}\ }\textbf {\bibinfo {volume} {32}},\ \bibinfo {pages} {215020} (\bibinfo {year} {2015})}\BibitemShut {NoStop}%
\bibitem [{\citenamefont {Ovalle}(2016)}]{ovalle2016extending}%
  \BibitemOpen
  \bibfield  {author} {\bibinfo {author} {\bibfnamefont {J.}~\bibnamefont {Ovalle}},\ }\bibfield  {title} {\bibinfo {title} {Extending the geometric deformation: New black hole solutions},\ }in\ \href@noop {} {\emph {\bibinfo {booktitle} {Int. J. Mod. Phys. Conf. Ser.}}},\ Vol.~\bibinfo {volume} {41}\ (\bibinfo {organization} {World Scientific},\ \bibinfo {year} {2016})\ p.\ \bibinfo {pages} {1660132}\BibitemShut {NoStop}%
\bibitem [{\citenamefont {Ovalle}(2009)}]{ovalle2009nonuniform}%
  \BibitemOpen
  \bibfield  {author} {\bibinfo {author} {\bibfnamefont {J.}~\bibnamefont {Ovalle}},\ }\bibfield  {title} {\bibinfo {title} {Nonuniform braneworld stars: an exact solution},\ }\href@noop {} {\bibfield  {journal} {\bibinfo  {journal} {Int. J. Mod. Phys. D}\ }\textbf {\bibinfo {volume} {18}},\ \bibinfo {pages} {837} (\bibinfo {year} {2009})}\BibitemShut {NoStop}%
\bibitem [{\citenamefont {Ovalle}(2010)}]{ovalle2010schwarzschild}%
  \BibitemOpen
  \bibfield  {author} {\bibinfo {author} {\bibfnamefont {J.}~\bibnamefont {Ovalle}},\ }\bibfield  {title} {\bibinfo {title} {The schwarzschild's braneworld solution},\ }\href@noop {} {\bibfield  {journal} {\bibinfo  {journal} {Mod. Phys. Lett. A}\ }\textbf {\bibinfo {volume} {25}},\ \bibinfo {pages} {3323} (\bibinfo {year} {2010})}\BibitemShut {NoStop}%
\bibitem [{\citenamefont {Casadio}\ and\ \citenamefont {Ovalle}(2012)}]{casadio2012brane}%
  \BibitemOpen
  \bibfield  {author} {\bibinfo {author} {\bibfnamefont {R.}~\bibnamefont {Casadio}}\ and\ \bibinfo {author} {\bibfnamefont {J.}~\bibnamefont {Ovalle}},\ }\bibfield  {title} {\bibinfo {title} {Brane-world stars and (microscopic) black holes},\ }\href@noop {} {\bibfield  {journal} {\bibinfo  {journal} {Phys. Lett. B}\ }\textbf {\bibinfo {volume} {715}},\ \bibinfo {pages} {251} (\bibinfo {year} {2012})}\BibitemShut {NoStop}%
\bibitem [{\citenamefont {Ovalle}\ and\ \citenamefont {Linares}(2013)}]{ovalle2013tolman}%
  \BibitemOpen
  \bibfield  {author} {\bibinfo {author} {\bibfnamefont {J.}~\bibnamefont {Ovalle}}\ and\ \bibinfo {author} {\bibfnamefont {F.}~\bibnamefont {Linares}},\ }\bibfield  {title} {\bibinfo {title} {Tolman iv solution in the \textsc{R}andall-\textsc{S}undrum braneworld},\ }\href@noop {} {\bibfield  {journal} {\bibinfo  {journal} {Phys. Rev. D}\ }\textbf {\bibinfo {volume} {88}},\ \bibinfo {pages} {104026} (\bibinfo {year} {2013})}\BibitemShut {NoStop}%
\bibitem [{\citenamefont {Ovalle}\ \emph {et~al.}(2013)\citenamefont {Ovalle}, \citenamefont {Linares}, \citenamefont {Pasqua},\ and\ \citenamefont {Sotomayor}}]{ovalle2013role}%
  \BibitemOpen
  \bibfield  {author} {\bibinfo {author} {\bibfnamefont {J.}~\bibnamefont {Ovalle}}, \bibinfo {author} {\bibfnamefont {F.}~\bibnamefont {Linares}}, \bibinfo {author} {\bibfnamefont {A.}~\bibnamefont {Pasqua}},\ and\ \bibinfo {author} {\bibfnamefont {A.}~\bibnamefont {Sotomayor}},\ }\bibfield  {title} {\bibinfo {title} {The role of exterior \textsc{W}eyl fluids on compact stellar structures in \textsc{R}andall--\textsc{S}undrum gravity},\ }\href@noop {} {\bibfield  {journal} {\bibinfo  {journal} {Class. Quantum Grav.}\ }\textbf {\bibinfo {volume} {30}},\ \bibinfo {pages} {175019} (\bibinfo {year} {2013})}\BibitemShut {NoStop}%
\bibitem [{\citenamefont {Casadio}\ and\ \citenamefont {da~Rocha}(2016)}]{casadio2016stability}%
  \BibitemOpen
  \bibfield  {author} {\bibinfo {author} {\bibfnamefont {R.}~\bibnamefont {Casadio}}\ and\ \bibinfo {author} {\bibfnamefont {R.}~\bibnamefont {da~Rocha}},\ }\bibfield  {title} {\bibinfo {title} {Stability of the graviton bose--einstein condensate in the brane-world},\ }\href@noop {} {\bibfield  {journal} {\bibinfo  {journal} {Phys. Lett. B}\ }\textbf {\bibinfo {volume} {763}},\ \bibinfo {pages} {434} (\bibinfo {year} {2016})}\BibitemShut {NoStop}%
\bibitem [{\citenamefont {da~Rocha}(2017)}]{da2017dark}%
  \BibitemOpen
  \bibfield  {author} {\bibinfo {author} {\bibfnamefont {R.}~\bibnamefont {da~Rocha}},\ }\bibfield  {title} {\bibinfo {title} {Dark $\textsc{SU}(${N}$)$ glueball stars on fluid branes},\ }\href@noop {} {\bibfield  {journal} {\bibinfo  {journal} {Phys. Rev. D}\ }\textbf {\bibinfo {volume} {95}},\ \bibinfo {pages} {124017} (\bibinfo {year} {2017})}\BibitemShut {NoStop}%
\bibitem [{\citenamefont {Fernandes-Silva}\ and\ \citenamefont {Rocha}(2018)}]{fernandes2018gregory}%
  \BibitemOpen
  \bibfield  {author} {\bibinfo {author} {\bibfnamefont {A.}~\bibnamefont {Fernandes-Silva}}\ and\ \bibinfo {author} {\bibfnamefont {R.~d.}\ \bibnamefont {Rocha}},\ }\bibfield  {title} {\bibinfo {title} {Gregory--laflamme analysis of \textsc{MGD} black strings},\ }\href@noop {} {\bibfield  {journal} {\bibinfo  {journal} {Eur. Phys. J. C}\ }\textbf {\bibinfo {volume} {78}},\ \bibinfo {pages} {271} (\bibinfo {year} {2018})}\BibitemShut {NoStop}%
\bibitem [{\citenamefont {Contreras}\ and\ \citenamefont {Bargue{\~n}o}(2018{\natexlab{a}})}]{contreras2018minimal}%
  \BibitemOpen
  \bibfield  {author} {\bibinfo {author} {\bibfnamefont {E.}~\bibnamefont {Contreras}}\ and\ \bibinfo {author} {\bibfnamefont {P.}~\bibnamefont {Bargue{\~n}o}},\ }\bibfield  {title} {\bibinfo {title} {Minimal geometric deformation decoupling in 2+1 dimensional space--times},\ }\href@noop {} {\bibfield  {journal} {\bibinfo  {journal} {Eur. Phys. J.l C}\ }\textbf {\bibinfo {volume} {78}},\ \bibinfo {pages} {558} (\bibinfo {year} {2018}{\natexlab{a}})}\BibitemShut {NoStop}%
\bibitem [{\citenamefont {Panotopoulos}\ and\ \citenamefont {Rinc{\'o}n}(2018)}]{panotopoulos2018minimal}%
  \BibitemOpen
  \bibfield  {author} {\bibinfo {author} {\bibfnamefont {G.}~\bibnamefont {Panotopoulos}}\ and\ \bibinfo {author} {\bibfnamefont {{\'A}.}~\bibnamefont {Rinc{\'o}n}},\ }\bibfield  {title} {\bibinfo {title} {Minimal geometric deformation in a cloud of strings},\ }\href@noop {} {\bibfield  {journal} {\bibinfo  {journal} {Eur. Phys. J. C}\ }\textbf {\bibinfo {volume} {78}},\ \bibinfo {pages} {851} (\bibinfo {year} {2018})}\BibitemShut {NoStop}%
\bibitem [{\citenamefont {Contreras}\ and\ \citenamefont {Bargue{\~n}o}(2018{\natexlab{b}})}]{acontreras2018minimal}%
  \BibitemOpen
  \bibfield  {author} {\bibinfo {author} {\bibfnamefont {E.}~\bibnamefont {Contreras}}\ and\ \bibinfo {author} {\bibfnamefont {P.}~\bibnamefont {Bargue{\~n}o}},\ }\bibfield  {title} {\bibinfo {title} {Minimal geometric deformation in asymptotically (\textsc{A}-) d\textsc{S} space-times and the isotropic sector for a polytropic black hole},\ }\href@noop {} {\bibfield  {journal} {\bibinfo  {journal} {Eur. Phys. J. C}\ }\textbf {\bibinfo {volume} {78}},\ \bibinfo {pages} {985} (\bibinfo {year} {2018}{\natexlab{b}})}\BibitemShut {NoStop}%
\bibitem [{\citenamefont {Gabbanelli}\ \emph {et~al.}(2019)\citenamefont {Gabbanelli}, \citenamefont {Ovalle}, \citenamefont {Sotomayor}, \citenamefont {Stuchlik},\ and\ \citenamefont {Casadio}}]{gabbanelli2019causal}%
  \BibitemOpen
  \bibfield  {author} {\bibinfo {author} {\bibfnamefont {L.}~\bibnamefont {Gabbanelli}}, \bibinfo {author} {\bibfnamefont {J.}~\bibnamefont {Ovalle}}, \bibinfo {author} {\bibfnamefont {A.}~\bibnamefont {Sotomayor}}, \bibinfo {author} {\bibfnamefont {Z.}~\bibnamefont {Stuchlik}},\ and\ \bibinfo {author} {\bibfnamefont {R.}~\bibnamefont {Casadio}},\ }\bibfield  {title} {\bibinfo {title} {A causal schwarzschild-de sitter interior solution by gravitational decoupling},\ }\href@noop {} {\bibfield  {journal} {\bibinfo  {journal} {Eur. Phys. J. C}\ }\textbf {\bibinfo {volume} {79}},\ \bibinfo {pages} {486} (\bibinfo {year} {2019})}\BibitemShut {NoStop}%
\bibitem [{\citenamefont {Hensh}\ and\ \citenamefont {Stuchl{\'\i}k}(2019)}]{hensh2019anisotropic}%
  \BibitemOpen
  \bibfield  {author} {\bibinfo {author} {\bibfnamefont {S.}~\bibnamefont {Hensh}}\ and\ \bibinfo {author} {\bibfnamefont {Z.}~\bibnamefont {Stuchl{\'\i}k}},\ }\bibfield  {title} {\bibinfo {title} {Anisotropic tolman vii solution by gravitational decoupling},\ }\href@noop {} {\bibfield  {journal} {\bibinfo  {journal} {Eur. Phys. J. C}\ }\textbf {\bibinfo {volume} {79}},\ \bibinfo {pages} {834} (\bibinfo {year} {2019})}\BibitemShut {NoStop}%
\bibitem [{\citenamefont {Torres-S{\'a}nchez}\ and\ \citenamefont {Contreras}(2019)}]{torres2019anisotropic}%
  \BibitemOpen
  \bibfield  {author} {\bibinfo {author} {\bibfnamefont {V.}~\bibnamefont {Torres-S{\'a}nchez}}\ and\ \bibinfo {author} {\bibfnamefont {E.}~\bibnamefont {Contreras}},\ }\bibfield  {title} {\bibinfo {title} {Anisotropic neutron stars by gravitational decoupling},\ }\href@noop {} {\bibfield  {journal} {\bibinfo  {journal} {Eur. Phys. J. C}\ }\textbf {\bibinfo {volume} {79}},\ \bibinfo {pages} {829} (\bibinfo {year} {2019})}\BibitemShut {NoStop}%
\bibitem [{\citenamefont {Ovalle}(2019)}]{ovalle2019decoupling}%
  \BibitemOpen
  \bibfield  {author} {\bibinfo {author} {\bibfnamefont {J.}~\bibnamefont {Ovalle}},\ }\bibfield  {title} {\bibinfo {title} {Decoupling gravitational sources in general relativity: The extended case},\ }\href@noop {} {\bibfield  {journal} {\bibinfo  {journal} {Phys. Lett. B}\ }\textbf {\bibinfo {volume} {788}},\ \bibinfo {pages} {213} (\bibinfo {year} {2019})}\BibitemShut {NoStop}%
\bibitem [{\citenamefont {Contreras}(2018)}]{zcontreras2018minimal}%
  \BibitemOpen
  \bibfield  {author} {\bibinfo {author} {\bibfnamefont {E.}~\bibnamefont {Contreras}},\ }\bibfield  {title} {\bibinfo {title} {Minimal geometric deformation: the inverse problem},\ }\href@noop {} {\bibfield  {journal} {\bibinfo  {journal} {Eur. Phys. J. C}\ }\textbf {\bibinfo {volume} {78}},\ \bibinfo {pages} {678} (\bibinfo {year} {2018})}\BibitemShut {NoStop}%
\bibitem [{\citenamefont {Ovalle}\ \emph {et~al.}(2018{\natexlab{b}})\citenamefont {Ovalle}, \citenamefont {Casadio}, \citenamefont {Rocha}, \citenamefont {Sotomayor},\ and\ \citenamefont {Stuchl{\'\i}k}}]{ovalle2018black}%
  \BibitemOpen
  \bibfield  {author} {\bibinfo {author} {\bibfnamefont {J.}~\bibnamefont {Ovalle}}, \bibinfo {author} {\bibfnamefont {R.}~\bibnamefont {Casadio}}, \bibinfo {author} {\bibfnamefont {R.~d.}\ \bibnamefont {Rocha}}, \bibinfo {author} {\bibfnamefont {A.}~\bibnamefont {Sotomayor}},\ and\ \bibinfo {author} {\bibfnamefont {Z.}~\bibnamefont {Stuchl{\'\i}k}},\ }\bibfield  {title} {\bibinfo {title} {Black holes by gravitational decoupling},\ }\href@noop {} {\bibfield  {journal} {\bibinfo  {journal} {Eur. Phys. J. C}\ }\textbf {\bibinfo {volume} {78}},\ \bibinfo {pages} {960} (\bibinfo {year} {2018}{\natexlab{b}})}\BibitemShut {NoStop}%
\bibitem [{\citenamefont {Tello-Ortiz}(2020)}]{tello2020minimally}%
  \BibitemOpen
  \bibfield  {author} {\bibinfo {author} {\bibfnamefont {F.}~\bibnamefont {Tello-Ortiz}},\ }\bibfield  {title} {\bibinfo {title} {Minimally deformed anisotropic dark stars in the framework of gravitational decoupling},\ }\href@noop {} {\bibfield  {journal} {\bibinfo  {journal} {Eur. Phys. J. C}\ }\textbf {\bibinfo {volume} {80}},\ \bibinfo {pages} {413} (\bibinfo {year} {2020})}\BibitemShut {NoStop}%
\bibitem [{\citenamefont {Cede{\~n}o}\ and\ \citenamefont {Contreras}(2020)}]{cedeno2020gravitational}%
  \BibitemOpen
  \bibfield  {author} {\bibinfo {author} {\bibfnamefont {F.~X.~L.}\ \bibnamefont {Cede{\~n}o}}\ and\ \bibinfo {author} {\bibfnamefont {E.}~\bibnamefont {Contreras}},\ }\bibfield  {title} {\bibinfo {title} {Gravitational decoupling in cosmology},\ }\href@noop {} {\bibfield  {journal} {\bibinfo  {journal} {Phys. Dark Universe}\ }\textbf {\bibinfo {volume} {28}},\ \bibinfo {pages} {100543} (\bibinfo {year} {2020})}\BibitemShut {NoStop}%
\bibitem [{\citenamefont {Maurya}\ \emph {et~al.}(2024)\citenamefont {Maurya}, \citenamefont {Singh}, \citenamefont {Aziz}, \citenamefont {Ray},\ and\ \citenamefont {Mustafa}}]{maurya2024compact}%
  \BibitemOpen
  \bibfield  {author} {\bibinfo {author} {\bibfnamefont {S.~K.}\ \bibnamefont {Maurya}}, \bibinfo {author} {\bibfnamefont {K.~N.}\ \bibnamefont {Singh}}, \bibinfo {author} {\bibfnamefont {A.}~\bibnamefont {Aziz}}, \bibinfo {author} {\bibfnamefont {S.}~\bibnamefont {Ray}},\ and\ \bibinfo {author} {\bibfnamefont {G.}~\bibnamefont {Mustafa}},\ }\bibfield  {title} {\bibinfo {title} {Compact stars with dark matter induced anisotropy in complexity-free background and effect of dark matter on gw echoes},\ }\href@noop {} {\bibfield  {journal} {\bibinfo  {journal} {Mon. Not. R. Astron. Soc.}\ }\textbf {\bibinfo {volume} {527}},\ \bibinfo {pages} {5192} (\bibinfo {year} {2024})}\BibitemShut {NoStop}%
\bibitem [{\citenamefont {Maurya}\ \emph {et~al.}(2023)\citenamefont {Maurya}, \citenamefont {Errehymy}, \citenamefont {Singh}, \citenamefont {Al-Harbi}, \citenamefont {Nisar}, \citenamefont {Abdel-Aty} \emph {et~al.}}]{maurya2023minimally}%
  \BibitemOpen
  \bibfield  {author} {\bibinfo {author} {\bibfnamefont {S.}~\bibnamefont {Maurya}}, \bibinfo {author} {\bibfnamefont {A.}~\bibnamefont {Errehymy}}, \bibinfo {author} {\bibfnamefont {K.}~\bibnamefont {Singh}}, \bibinfo {author} {\bibfnamefont {N.}~\bibnamefont {Al-Harbi}}, \bibinfo {author} {\bibfnamefont {K.~S.}\ \bibnamefont {Nisar}}, \bibinfo {author} {\bibfnamefont {A.-H.}\ \bibnamefont {Abdel-Aty}}, \emph {et~al.},\ }\bibfield  {title} {\bibinfo {title} {Minimally deformed anisotropic stars in dark matter halos under egb-action},\ }\href@noop {} {\bibfield  {journal} {\bibinfo  {journal} {Eur. Phys. J. C}\ }\textbf {\bibinfo {volume} {83}},\ \bibinfo {pages} {968} (\bibinfo {year} {2023})}\BibitemShut {NoStop}%
\bibitem [{\citenamefont {Yousaf}\ \emph {et~al.}(2023)\citenamefont {Yousaf}, \citenamefont {Adeel}, \citenamefont {Khan},\ and\ \citenamefont {Bhatti}}]{yousaf2023generating}%
  \BibitemOpen
  \bibfield  {author} {\bibinfo {author} {\bibfnamefont {Z.}~\bibnamefont {Yousaf}}, \bibinfo {author} {\bibfnamefont {A.}~\bibnamefont {Adeel}}, \bibinfo {author} {\bibfnamefont {S.}~\bibnamefont {Khan}},\ and\ \bibinfo {author} {\bibfnamefont {M.~Z.}\ \bibnamefont {Bhatti}},\ }\bibfield  {title} {\bibinfo {title} {Generating fuzzy dark matter droplets},\ }\href@noop {} {\bibfield  {journal} {\bibinfo  {journal} {Chinese J. Phys.}\ }\textbf {\bibinfo {volume} {88}},\ \bibinfo {pages} {406} (\bibinfo {year} {2023})}\BibitemShut {NoStop}%
\bibitem [{\citenamefont {Ivanytskyi}\ \emph {et~al.}(2020)\citenamefont {Ivanytskyi}, \citenamefont {Sagun},\ and\ \citenamefont {Lopes}}]{ivanytskyi2020neutron}%
  \BibitemOpen
  \bibfield  {author} {\bibinfo {author} {\bibfnamefont {O.}~\bibnamefont {Ivanytskyi}}, \bibinfo {author} {\bibfnamefont {V.}~\bibnamefont {Sagun}},\ and\ \bibinfo {author} {\bibfnamefont {I.}~\bibnamefont {Lopes}},\ }\bibfield  {title} {\bibinfo {title} {Neutron stars: New constraints on asymmetric dark matter},\ }\href@noop {} {\bibfield  {journal} {\bibinfo  {journal} {Phys. Rev. D}\ }\textbf {\bibinfo {volume} {102}},\ \bibinfo {pages} {063028} (\bibinfo {year} {2020})}\BibitemShut {NoStop}%
\bibitem [{\citenamefont {Freese}\ \emph {et~al.}(2008)\citenamefont {Freese}, \citenamefont {Bodenheimer}, \citenamefont {Spolyar},\ and\ \citenamefont {Gondolo}}]{freese2008stellar}%
  \BibitemOpen
  \bibfield  {author} {\bibinfo {author} {\bibfnamefont {K.}~\bibnamefont {Freese}}, \bibinfo {author} {\bibfnamefont {P.}~\bibnamefont {Bodenheimer}}, \bibinfo {author} {\bibfnamefont {D.}~\bibnamefont {Spolyar}},\ and\ \bibinfo {author} {\bibfnamefont {P.}~\bibnamefont {Gondolo}},\ }\bibfield  {title} {\bibinfo {title} {Stellar structure of dark stars: A first phase of stellar evolution resulting from dark matter annihilation},\ }\href@noop {} {\bibfield  {journal} {\bibinfo  {journal} {Astrophys. J.}\ }\textbf {\bibinfo {volume} {685}},\ \bibinfo {pages} {L101} (\bibinfo {year} {2008})}\BibitemShut {NoStop}%
\bibitem [{\citenamefont {de~Lavallaz}\ and\ \citenamefont {Fairbairn}(2010)}]{de2010neutron}%
  \BibitemOpen
  \bibfield  {author} {\bibinfo {author} {\bibfnamefont {A.}~\bibnamefont {de~Lavallaz}}\ and\ \bibinfo {author} {\bibfnamefont {M.}~\bibnamefont {Fairbairn}},\ }\bibfield  {title} {\bibinfo {title} {Neutron stars as dark matter probes},\ }\href@noop {} {\bibfield  {journal} {\bibinfo  {journal} {Phys. Rev. D}\ }\textbf {\bibinfo {volume} {81}},\ \bibinfo {pages} {123521} (\bibinfo {year} {2010})}\BibitemShut {NoStop}%
\bibitem [{\citenamefont {Raen}\ \emph {et~al.}(2021)\citenamefont {Raen}, \citenamefont {Mart{\'\i}nez-Rodr{\'\i}guez}, \citenamefont {Hurst}, \citenamefont {Zentner}, \citenamefont {Badenes},\ and\ \citenamefont {Tao}}]{raen2021effects}%
  \BibitemOpen
  \bibfield  {author} {\bibinfo {author} {\bibfnamefont {T.~J.}\ \bibnamefont {Raen}}, \bibinfo {author} {\bibfnamefont {H.}~\bibnamefont {Mart{\'\i}nez-Rodr{\'\i}guez}}, \bibinfo {author} {\bibfnamefont {T.~J.}\ \bibnamefont {Hurst}}, \bibinfo {author} {\bibfnamefont {A.~R.}\ \bibnamefont {Zentner}}, \bibinfo {author} {\bibfnamefont {C.}~\bibnamefont {Badenes}},\ and\ \bibinfo {author} {\bibfnamefont {R.}~\bibnamefont {Tao}},\ }\bibfield  {title} {\bibinfo {title} {The effects of asymmetric dark matter on stellar evolution--\textsc{I}. spin-dependent scattering},\ }\href@noop {} {\bibfield  {journal} {\bibinfo  {journal} {Mon. Not. R. Astron. Soc.}\ }\textbf {\bibinfo {volume} {503}},\ \bibinfo {pages} {5611} (\bibinfo {year} {2021})}\BibitemShut {NoStop}%
\bibitem [{\citenamefont {Chakraborty}\ and\ \citenamefont {SenGupta}(2018)}]{chakraborty2018packing}%
  \BibitemOpen
  \bibfield  {author} {\bibinfo {author} {\bibfnamefont {S.}~\bibnamefont {Chakraborty}}\ and\ \bibinfo {author} {\bibfnamefont {S.}~\bibnamefont {SenGupta}},\ }\bibfield  {title} {\bibinfo {title} {Packing extra mass in compact stellar structures: An interplay between \textsc{K}alb-\textsc{R}amond field and extra dimensions},\ }\href@noop {} {\bibfield  {journal} {\bibinfo  {journal} {J. Cosmol. Astropart. Phys.}\ }\textbf {\bibinfo {volume} {2018}}\bibinfo  {number} { (05)},\ \bibinfo {pages} {032}}\BibitemShut {NoStop}%
\bibitem [{\citenamefont {Sharif}\ and\ \citenamefont {Naseer}(2023{\natexlab{a}})}]{sharif2023effects}%
  \BibitemOpen
\bibfield  {number} {  }\bibfield  {author} {\bibinfo {author} {\bibfnamefont {M.}~\bibnamefont {Sharif}}\ and\ \bibinfo {author} {\bibfnamefont {T.}~\bibnamefont {Naseer}},\ }\bibfield  {title} {\bibinfo {title} {Effects of charge and gravitational decoupling on complexity and isotropization of anisotropic models},\ }\href@noop {} {\bibfield  {journal} {\bibinfo  {journal} {Phys. Dark Universe}\ }\textbf {\bibinfo {volume} {42}},\ \bibinfo {pages} {101324} (\bibinfo {year} {2023}{\natexlab{a}})}\BibitemShut {NoStop}%
\bibitem [{\citenamefont {Sharif}\ and\ \citenamefont {Naseer}(2023{\natexlab{b}})}]{sharif2023charge}%
  \BibitemOpen
  \bibfield  {author} {\bibinfo {author} {\bibfnamefont {M.}~\bibnamefont {Sharif}}\ and\ \bibinfo {author} {\bibfnamefont {T.}~\bibnamefont {Naseer}},\ }\bibfield  {title} {\bibinfo {title} {Charge effect on isotropization and complexity of extended decoupled anisotropic stellar models},\ }\href@noop {} {\bibfield  {journal} {\bibinfo  {journal} {Chinese J. Phys.}\ }\textbf {\bibinfo {volume} {86}},\ \bibinfo {pages} {596} (\bibinfo {year} {2023}{\natexlab{b}})}\BibitemShut {NoStop}%
\bibitem [{\citenamefont {Sharif}\ and\ \citenamefont {Naseer}(2023{\natexlab{c}})}]{sharif2023effect}%
  \BibitemOpen
  \bibfield  {author} {\bibinfo {author} {\bibfnamefont {M.}~\bibnamefont {Sharif}}\ and\ \bibinfo {author} {\bibfnamefont {T.}~\bibnamefont {Naseer}},\ }\bibfield  {title} {\bibinfo {title} {Effect of extended gravitational decoupling on isotropization and complexity in theory},\ }\href@noop {} {\bibfield  {journal} {\bibinfo  {journal} {Class. Quantum Grav.}\ }\textbf {\bibinfo {volume} {40}},\ \bibinfo {pages} {035009} (\bibinfo {year} {2023}{\natexlab{c}})}\BibitemShut {NoStop}%
\bibitem [{\citenamefont {Yousaf}\ \emph {et~al.}(2024{\natexlab{b}})\citenamefont {Yousaf}, \citenamefont {Bamba}, \citenamefont {Almutairi}, \citenamefont {Khan},\ and\ \citenamefont {Bhatti}}]{yousaf2024role}%
  \BibitemOpen
  \bibfield  {author} {\bibinfo {author} {\bibfnamefont {Z.}~\bibnamefont {Yousaf}}, \bibinfo {author} {\bibfnamefont {K.}~\bibnamefont {Bamba}}, \bibinfo {author} {\bibfnamefont {B.}~\bibnamefont {Almutairi}}, \bibinfo {author} {\bibfnamefont {S.}~\bibnamefont {Khan}},\ and\ \bibinfo {author} {\bibfnamefont {M.~Z.}\ \bibnamefont {Bhatti}},\ }\bibfield  {title} {\bibinfo {title} {Role of complexity on the minimal deformation of black holes},\ }\href@noop {} {\bibfield  {journal} {\bibinfo  {journal} {Class. Quantum Grav.}\ }\textbf {\bibinfo {volume} {41}},\ \bibinfo {pages} {175001} (\bibinfo {year} {2024}{\natexlab{b}})}\BibitemShut {NoStop}%
\bibitem [{\citenamefont {Einasto}(1965)}]{einasto1965construction}%
  \BibitemOpen
  \bibfield  {author} {\bibinfo {author} {\bibfnamefont {J.}~\bibnamefont {Einasto}},\ }\bibfield  {title} {\bibinfo {title} {On the construction of a composite model for the galaxy and on the determination of the system of galactic parameters},\ }\href@noop {} {\bibfield  {journal} {\bibinfo  {journal} {Trudy Astrofizicheskogo Instituta Alma-Ata}\ }\textbf {\bibinfo {volume} {5}},\ \bibinfo {pages} {87} (\bibinfo {year} {1965})}\BibitemShut {NoStop}%
\bibitem [{\citenamefont {Retana-Montenegro}\ \emph {et~al.}(2012)\citenamefont {Retana-Montenegro}, \citenamefont {Van~Hese}, \citenamefont {Gentile}, \citenamefont {Baes},\ and\ \citenamefont {Frutos-Alfaro}}]{retana2012analytical}%
  \BibitemOpen
  \bibfield  {author} {\bibinfo {author} {\bibfnamefont {E.}~\bibnamefont {Retana-Montenegro}}, \bibinfo {author} {\bibfnamefont {E.}~\bibnamefont {Van~Hese}}, \bibinfo {author} {\bibfnamefont {G.}~\bibnamefont {Gentile}}, \bibinfo {author} {\bibfnamefont {M.}~\bibnamefont {Baes}},\ and\ \bibinfo {author} {\bibfnamefont {F.}~\bibnamefont {Frutos-Alfaro}},\ }\bibfield  {title} {\bibinfo {title} {Analytical properties of \textsc{E}inasto dark matter haloes},\ }\href@noop {} {\bibfield  {journal} {\bibinfo  {journal} {Astron. Astrophys.}\ }\textbf {\bibinfo {volume} {540}},\ \bibinfo {pages} {A70} (\bibinfo {year} {2012})}\BibitemShut {NoStop}%
\bibitem [{\citenamefont {Einasto}(1969{\natexlab{a}})}]{einasto1969andromeda}%
  \BibitemOpen
  \bibfield  {author} {\bibinfo {author} {\bibfnamefont {J.}~\bibnamefont {Einasto}},\ }\bibfield  {title} {\bibinfo {title} {The \textsc{A}ndromeda galaxy \textsc{M} 31: I. a preliminary model},\ }\href@noop {} {\bibfield  {journal} {\bibinfo  {journal} {Astrophysics}\ }\textbf {\bibinfo {volume} {5}},\ \bibinfo {pages} {67} (\bibinfo {year} {1969}{\natexlab{a}})}\BibitemShut {NoStop}%
\bibitem [{\citenamefont {Einasto}(1969{\natexlab{b}})}]{einasto1969galactic}%
  \BibitemOpen
  \bibfield  {author} {\bibinfo {author} {\bibfnamefont {J.}~\bibnamefont {Einasto}},\ }\bibfield  {title} {\bibinfo {title} {On galactic descriptive functions},\ }\href@noop {} {\bibfield  {journal} {\bibinfo  {journal} {Astron. Nachr.}\ }\textbf {\bibinfo {volume} {291}},\ \bibinfo {pages} {97} (\bibinfo {year} {1969}{\natexlab{b}})}\BibitemShut {NoStop}%
\bibitem [{\citenamefont {Navarro}\ \emph {et~al.}(2004)\citenamefont {Navarro} \emph {et~al.}}]{navarro2004inner}%
  \BibitemOpen
  \bibfield  {author} {\bibinfo {author} {\bibfnamefont {J.~F.}\ \bibnamefont {Navarro}} \emph {et~al.},\ }\bibfield  {title} {\bibinfo {title} {The inner structure of ${\Lambda}$\textsc{CDM} haloes--\textsc{III}. universality and asymptotic slopes},\ }\href@noop {} {\bibfield  {journal} {\bibinfo  {journal} {Mon. Not. R. Astron. Soc.}\ }\textbf {\bibinfo {volume} {349}},\ \bibinfo {pages} {1039} (\bibinfo {year} {2004})}\BibitemShut {NoStop}%
\bibitem [{\citenamefont {Hayashi}\ and\ \citenamefont {White}(2008)}]{hayashi2008understanding}%
  \BibitemOpen
  \bibfield  {author} {\bibinfo {author} {\bibfnamefont {E.}~\bibnamefont {Hayashi}}\ and\ \bibinfo {author} {\bibfnamefont {S.~D.~M.}\ \bibnamefont {White}},\ }\bibfield  {title} {\bibinfo {title} {Understanding the halo-mass and galaxy-mass cross-correlation functions},\ }\href@noop {} {\bibfield  {journal} {\bibinfo  {journal} {Mon. Not. R. Astron. Soc.}\ }\textbf {\bibinfo {volume} {388}} (\bibinfo {year} {2008})}\BibitemShut {NoStop}%
\bibitem [{\citenamefont {Springel}\ \emph {et~al.}(2005)\citenamefont {Springel}, \citenamefont {White}, \citenamefont {Jenkins} \emph {et~al.}}]{springel2005simulations}%
  \BibitemOpen
  \bibfield  {author} {\bibinfo {author} {\bibfnamefont {V.}~\bibnamefont {Springel}}, \bibinfo {author} {\bibfnamefont {S.~D.~M.}\ \bibnamefont {White}}, \bibinfo {author} {\bibfnamefont {A.}~\bibnamefont {Jenkins}}, \emph {et~al.},\ }\bibfield  {title} {\bibinfo {title} {Simulations of the formation, evolution and clustering of galaxies and quasars},\ }\href@noop {} {\bibfield  {journal} {\bibinfo  {journal} {nature}\ }\textbf {\bibinfo {volume} {435}},\ \bibinfo {pages} {629} (\bibinfo {year} {2005})}\BibitemShut {NoStop}%
\bibitem [{\citenamefont {Gao}\ \emph {et~al.}(2008)\citenamefont {Gao} \emph {et~al.}}]{gao2008redshift}%
  \BibitemOpen
  \bibfield  {author} {\bibinfo {author} {\bibfnamefont {L.}~\bibnamefont {Gao}} \emph {et~al.},\ }\bibfield  {title} {\bibinfo {title} {The redshift dependence of the structure of massive ${\Lambda}$ cold dark matter haloes},\ }\href@noop {} {\bibfield  {journal} {\bibinfo  {journal} {Mon. Not. R. Astron. Soc.}\ }\textbf {\bibinfo {volume} {387}},\ \bibinfo {pages} {536} (\bibinfo {year} {2008})}\BibitemShut {NoStop}%
\bibitem [{\citenamefont {Ovalle}\ \emph {et~al.}(2022)\citenamefont {Ovalle}, \citenamefont {Contreras},\ and\ \citenamefont {Stuchlik}}]{ovalle2022energy}%
  \BibitemOpen
  \bibfield  {author} {\bibinfo {author} {\bibfnamefont {J.}~\bibnamefont {Ovalle}}, \bibinfo {author} {\bibfnamefont {E.}~\bibnamefont {Contreras}},\ and\ \bibinfo {author} {\bibfnamefont {Z.}~\bibnamefont {Stuchlik}},\ }\bibfield  {title} {\bibinfo {title} {Energy exchange between relativistic fluids: the polytropic case},\ }\href@noop {} {\bibfield  {journal} {\bibinfo  {journal} {Eur. Phys. J. C}\ }\textbf {\bibinfo {volume} {82}},\ \bibinfo {pages} {211} (\bibinfo {year} {2022})}\BibitemShut {NoStop}%
\bibitem [{\citenamefont {Contreras}\ and\ \citenamefont {Stuchlik}(2022)}]{contreras2022energy}%
  \BibitemOpen
  \bibfield  {author} {\bibinfo {author} {\bibfnamefont {E.}~\bibnamefont {Contreras}}\ and\ \bibinfo {author} {\bibfnamefont {Z.}~\bibnamefont {Stuchlik}},\ }\bibfield  {title} {\bibinfo {title} {Energy exchange between \textsc{T}olman \textsc{VII} and a polytropic fluid},\ }\href@noop {} {\bibfield  {journal} {\bibinfo  {journal} {Eur. Phys. J. C}\ }\textbf {\bibinfo {volume} {82}},\ \bibinfo {pages} {365} (\bibinfo {year} {2022})}\BibitemShut {NoStop}%
\bibitem [{\citenamefont {Khan}\ and\ \citenamefont {Yousaf}(2024)}]{khan2024complexity}%
  \BibitemOpen
  \bibfield  {author} {\bibinfo {author} {\bibfnamefont {S.}~\bibnamefont {Khan}}\ and\ \bibinfo {author} {\bibfnamefont {Z.}~\bibnamefont {Yousaf}},\ }\bibfield  {title} {\bibinfo {title} {Complexity-free charged anisotropic \textsc{F}inch-\textsc{S}kea model satisfying \textsc{K}armarkar condition},\ }\href@noop {} {\bibfield  {journal} {\bibinfo  {journal} {Phys. Scr.}\ }\textbf {\bibinfo {volume} {99}},\ \bibinfo {pages} {055303} (\bibinfo {year} {2024})}\BibitemShut {NoStop}%
\bibitem [{\citenamefont {Gabbanelli}\ \emph {et~al.}(2018)\citenamefont {Gabbanelli}, \citenamefont {Rinc{\'o}n},\ and\ \citenamefont {Rubio}}]{gabbanelli2018gravitational}%
  \BibitemOpen
  \bibfield  {author} {\bibinfo {author} {\bibfnamefont {L.}~\bibnamefont {Gabbanelli}}, \bibinfo {author} {\bibfnamefont {{\'A}.}~\bibnamefont {Rinc{\'o}n}},\ and\ \bibinfo {author} {\bibfnamefont {C.}~\bibnamefont {Rubio}},\ }\bibfield  {title} {\bibinfo {title} {Gravitational decoupled anisotropies in compact stars},\ }\href@noop {} {\bibfield  {journal} {\bibinfo  {journal} {Eur. Phys. J. C}\ }\textbf {\bibinfo {volume} {78}},\ \bibinfo {pages} {370} (\bibinfo {year} {2018})}\BibitemShut {NoStop}%
\bibitem [{\citenamefont {Morales}\ and\ \citenamefont {Tello-Ortiz}(2018{\natexlab{a}})}]{morales2018charged}%
  \BibitemOpen
  \bibfield  {author} {\bibinfo {author} {\bibfnamefont {E.}~\bibnamefont {Morales}}\ and\ \bibinfo {author} {\bibfnamefont {F.}~\bibnamefont {Tello-Ortiz}},\ }\bibfield  {title} {\bibinfo {title} {Charged anisotropic compact objects by gravitational decoupling},\ }\href@noop {} {\bibfield  {journal} {\bibinfo  {journal} {Eur. Phys. J. C}\ }\textbf {\bibinfo {volume} {78}},\ \bibinfo {pages} {618} (\bibinfo {year} {2018}{\natexlab{a}})}\BibitemShut {NoStop}%
\bibitem [{\citenamefont {Estrada}\ and\ \citenamefont {Prado}(2019)}]{estrada2019gravitational}%
  \BibitemOpen
  \bibfield  {author} {\bibinfo {author} {\bibfnamefont {M.}~\bibnamefont {Estrada}}\ and\ \bibinfo {author} {\bibfnamefont {R.}~\bibnamefont {Prado}},\ }\bibfield  {title} {\bibinfo {title} {The gravitational decoupling method: the higher-dimensional case to find new analytic solutions},\ }\href@noop {} {\bibfield  {journal} {\bibinfo  {journal} {Eur. Phys. J. Plus}\ }\textbf {\bibinfo {volume} {134}},\ \bibinfo {pages} {168} (\bibinfo {year} {2019})}\BibitemShut {NoStop}%
\bibitem [{\citenamefont {Maurya}\ and\ \citenamefont {Tello-Ortiz}(2019)}]{maurya2019generalized}%
  \BibitemOpen
  \bibfield  {author} {\bibinfo {author} {\bibfnamefont {S.}~\bibnamefont {Maurya}}\ and\ \bibinfo {author} {\bibfnamefont {F.}~\bibnamefont {Tello-Ortiz}},\ }\bibfield  {title} {\bibinfo {title} {Generalized relativistic anisotropic compact star models by gravitational decoupling},\ }\href@noop {} {\bibfield  {journal} {\bibinfo  {journal} {Eur. Phys. J. C}\ }\textbf {\bibinfo {volume} {79}},\ \bibinfo {pages} {85} (\bibinfo {year} {2019})}\BibitemShut {NoStop}%
\bibitem [{\citenamefont {Morales}\ and\ \citenamefont {Tello-Ortiz}(2018{\natexlab{b}})}]{morales2018compact}%
  \BibitemOpen
  \bibfield  {author} {\bibinfo {author} {\bibfnamefont {E.}~\bibnamefont {Morales}}\ and\ \bibinfo {author} {\bibfnamefont {F.}~\bibnamefont {Tello-Ortiz}},\ }\bibfield  {title} {\bibinfo {title} {Compact anisotropic models in general relativity by gravitational decoupling},\ }\href@noop {} {\bibfield  {journal} {\bibinfo  {journal} {Eur. Phys. J. C}\ }\textbf {\bibinfo {volume} {78}},\ \bibinfo {pages} {841} (\bibinfo {year} {2018}{\natexlab{b}})}\BibitemShut {NoStop}%
\bibitem [{\citenamefont {Abell{\'a}n}\ \emph {et~al.}(2020)\citenamefont {Abell{\'a}n}, \citenamefont {Torres-S{\'a}nchez}, \citenamefont {Fuenmayor},\ and\ \citenamefont {Contreras}}]{abellan2020regularity}%
  \BibitemOpen
  \bibfield  {author} {\bibinfo {author} {\bibfnamefont {G.}~\bibnamefont {Abell{\'a}n}}, \bibinfo {author} {\bibfnamefont {V.}~\bibnamefont {Torres-S{\'a}nchez}}, \bibinfo {author} {\bibfnamefont {E.}~\bibnamefont {Fuenmayor}},\ and\ \bibinfo {author} {\bibfnamefont {E.}~\bibnamefont {Contreras}},\ }\bibfield  {title} {\bibinfo {title} {Regularity condition on the anisotropy induced by gravitational decoupling in the framework of mgd},\ }\href@noop {} {\bibfield  {journal} {\bibinfo  {journal} {Eur. Phys. J. C}\ }\textbf {\bibinfo {volume} {80}},\ \bibinfo {pages} {177} (\bibinfo {year} {2020})}\BibitemShut {NoStop}%
\bibitem [{\citenamefont {Tello-Ortiz}\ \emph {et~al.}(2020)\citenamefont {Tello-Ortiz}, \citenamefont {Maurya},\ and\ \citenamefont {Gomez-Leyton}}]{tello2020class}%
  \BibitemOpen
  \bibfield  {author} {\bibinfo {author} {\bibfnamefont {F.}~\bibnamefont {Tello-Ortiz}}, \bibinfo {author} {\bibfnamefont {S.~K.}\ \bibnamefont {Maurya}},\ and\ \bibinfo {author} {\bibfnamefont {Y.}~\bibnamefont {Gomez-Leyton}},\ }\bibfield  {title} {\bibinfo {title} {Class \textsc{I} approach as \textsc{MGD} generator},\ }\href@noop {} {\bibfield  {journal} {\bibinfo  {journal} {Eur. Phys. J. C}\ }\textbf {\bibinfo {volume} {80}},\ \bibinfo {pages} {324} (\bibinfo {year} {2020})}\BibitemShut {NoStop}%
\bibitem [{\citenamefont {Maurya}\ \emph {et~al.}(2020)\citenamefont {Maurya}, \citenamefont {Tello-Ortiz},\ and\ \citenamefont {Jasim}}]{maurya2020egd}%
  \BibitemOpen
  \bibfield  {author} {\bibinfo {author} {\bibfnamefont {S.}~\bibnamefont {Maurya}}, \bibinfo {author} {\bibfnamefont {F.}~\bibnamefont {Tello-Ortiz}},\ and\ \bibinfo {author} {\bibfnamefont {M.}~\bibnamefont {Jasim}},\ }\bibfield  {title} {\bibinfo {title} {An egd model in the background of embedding class \textsc{I} space--time},\ }\href@noop {} {\bibfield  {journal} {\bibinfo  {journal} {Eur. Phys. J. C}\ }\textbf {\bibinfo {volume} {80}},\ \bibinfo {pages} {918} (\bibinfo {year} {2020})}\BibitemShut {NoStop}%
\bibitem [{\citenamefont {Buchdahl}(1959)}]{buchdahl1959general}%
  \BibitemOpen
  \bibfield  {author} {\bibinfo {author} {\bibfnamefont {H.~A.}\ \bibnamefont {Buchdahl}},\ }\bibfield  {title} {\bibinfo {title} {General relativistic fluid spheres},\ }\href@noop {} {\bibfield  {journal} {\bibinfo  {journal} {Phys. Rev.}\ }\textbf {\bibinfo {volume} {116}},\ \bibinfo {pages} {1027} (\bibinfo {year} {1959})}\BibitemShut {NoStop}%
\bibitem [{\citenamefont {Singh}\ \emph {et~al.}(2019)\citenamefont {Singh}, \citenamefont {Maurya}, \citenamefont {Jasim},\ and\ \citenamefont {Rahaman}}]{singh2019minimally}%
  \BibitemOpen
  \bibfield  {author} {\bibinfo {author} {\bibfnamefont {K.~N.}\ \bibnamefont {Singh}}, \bibinfo {author} {\bibfnamefont {S.}~\bibnamefont {Maurya}}, \bibinfo {author} {\bibfnamefont {M.}~\bibnamefont {Jasim}},\ and\ \bibinfo {author} {\bibfnamefont {F.}~\bibnamefont {Rahaman}},\ }\bibfield  {title} {\bibinfo {title} {Minimally deformed anisotropic model of class one space-time by gravitational decoupling},\ }\href@noop {} {\bibfield  {journal} {\bibinfo  {journal} {Eur. Phys. J. C}\ }\textbf {\bibinfo {volume} {79}},\ \bibinfo {pages} {851} (\bibinfo {year} {2019})}\BibitemShut {NoStop}%
\bibitem [{\citenamefont {Batic}\ \emph {et~al.}(2021)\citenamefont {Batic}, \citenamefont {Abuhejleh},\ and\ \citenamefont {Nowakowski}}]{batic2021fuzzy}%
  \BibitemOpen
  \bibfield  {author} {\bibinfo {author} {\bibfnamefont {D.}~\bibnamefont {Batic}}, \bibinfo {author} {\bibfnamefont {D.~A.}\ \bibnamefont {Abuhejleh}},\ and\ \bibinfo {author} {\bibfnamefont {M.}~\bibnamefont {Nowakowski}},\ }\bibfield  {title} {\bibinfo {title} {Fuzzy dark matter black holes and droplets},\ }\href@noop {} {\bibfield  {journal} {\bibinfo  {journal} {Eur. Phys. J. C}\ }\textbf {\bibinfo {volume} {81}},\ \bibinfo {pages} {777} (\bibinfo {year} {2021})}\BibitemShut {NoStop}%
\bibitem [{\citenamefont {Yousaf}\ \emph {et~al.}(2024{\natexlab{c}})\citenamefont {Yousaf}, \citenamefont {Adeel}, \citenamefont {Khan},\ and\ \citenamefont {Bhatti}}]{yousaf2024generating}%
  \BibitemOpen
  \bibfield  {author} {\bibinfo {author} {\bibfnamefont {Z.}~\bibnamefont {Yousaf}}, \bibinfo {author} {\bibfnamefont {A.}~\bibnamefont {Adeel}}, \bibinfo {author} {\bibfnamefont {S.}~\bibnamefont {Khan}},\ and\ \bibinfo {author} {\bibfnamefont {M.}~\bibnamefont {Bhatti}},\ }\bibfield  {title} {\bibinfo {title} {Generating fuzzy dark matter droplets},\ }\href@noop {} {\bibfield  {journal} {\bibinfo  {journal} {Chin. J. Phys.}\ }\textbf {\bibinfo {volume} {88}},\ \bibinfo {pages} {406} (\bibinfo {year} {2024}{\natexlab{c}})}\BibitemShut {NoStop}%
\bibitem [{\citenamefont {Khan}\ \emph {et~al.}(2024)\citenamefont {Khan}, \citenamefont {Adeel},\ and\ \citenamefont {Yousaf}}]{khan2024structure}%
  \BibitemOpen
  \bibfield  {author} {\bibinfo {author} {\bibfnamefont {S.}~\bibnamefont {Khan}}, \bibinfo {author} {\bibfnamefont {A.}~\bibnamefont {Adeel}},\ and\ \bibinfo {author} {\bibfnamefont {Z.}~\bibnamefont {Yousaf}},\ }\bibfield  {title} {\bibinfo {title} {Structure of anisotropic fuzzy dark matter black holes},\ }\href@noop {} {\bibfield  {journal} {\bibinfo  {journal} {Eur. Phys. J. C}\ }\textbf {\bibinfo {volume} {84}},\ \bibinfo {pages} {572} (\bibinfo {year} {2024})}\BibitemShut {NoStop}%
\end{thebibliography}
\end{document}